\documentclass[10pt]{article}
\usepackage{ametsoc2col}
\usepackage{graphicx}
\usepackage[normalem]{ulem}
\usepackage{float}
\usepackage[labelsep=endash]{caption}
\newcommand{\myabstract}{We present a revised and extended total and spectral solar irradiance (SSI) reconstruction, which includes a wavelength-dependent uncertainty estimate, spanning the last three solar cycles using the SATIRE-S model. The SSI reconstruction covers wavelengths between 115 and 160\,000 nm and all dates between August 1974 and October 2009. This represents the first full-wavelength SATIRE-S reconstruction to cover the last three solar cycles without data gaps and with an uncertainty estimate. SATIRE-S is compared with the NRLSSI model and SORCE/SOLSTICE ultraviolet (UV) observations. SATIRE-S displays similar cycle behaviour to NRLSSI for wavelengths below 242 nm and almost twice the variability between 242 and 310 nm. During the decline of last solar cycle, between 2003 and 2008, SSI from SORCE/SOLSTICE version 12 and 10 typically displays more than three times the variability of SATIRE-S between 200 and 300 nm. All three datasets are used to model changes in stratospheric ozone within a 2D atmospheric model for a decline from high solar activity to solar minimum. The different flux changes result in different modelled ozone trends. Using NRLSSI leads to a decline in mesospheric ozone, while SATIRE-S and SORCE/SOLSTICE result in an increase. Recent publications have highlighted increases in mesospheric ozone when considering version 10 SORCE/SOLSTICE irradiances. The recalibrated SORCE/SOLSTICE version 12 irradiances result in a much smaller mesospheric ozone response than when using version 10 and now similar in magnitude to SATIRE-S. This shows that current knowledge of variations in spectral irradiance is not sufficient to warrant robust conclusions concerning the impact of solar variability on the atmosphere and climate.}
\begin{document}
%
%
\title{\textbf{\large{A new SATIRE-S spectral solar irradiance reconstruction for solar cycles 21--23 and its implications for stratospheric ozone}}}
%
%
\author{\textsc{William T. Ball,}
				\thanks{\textit{Corresponding author address:} 
				William T. Ball, Physics Department, Blackett Laboratory, 
				Imperial College London, SW7 2AZ, United Kingdom. 
				\newline{E-mail: william.ball@imperial.ac.uk}}\\
\textit{\footnotesize{Physics Department, Blackett Laboratory, Imperial College London, SW7 2AZ, UK}}\\
\and 
\centerline{\textsc{Natalie A. Krivova}}\\
\centerline{\textit{\footnotesize{Max-Planck-Institut f\"ur Sonnensystemforschung, 37077 G\"ottingen, Germany}}}\\
\centerline{\textsc{Yvonne C. Unruh}}\\
\centerline{\textit{\footnotesize{Physics Department, Blackett Laboratory, Imperial College London, SW7 2AZ, UK}}}\\
\centerline{\textsc{Joanna D. Haigh}}\\
\centerline{\textit{\footnotesize{Physics Department, Blackett Laboratory, Imperial College London, SW7 2AZ, UK}}}\\
\centerline{\textit{\footnotesize{Grantham Institute, Imperial College London, SW7 2AZ, UK}}}\\
\centerline{\textsc{Sami K. Solanki}}\\
\centerline{\textit{\footnotesize{Max-Planck-Institut f\"ur Sonnensystemforschung, 37077 G\"ottingen, Germany}}}\\
\centerline{\textit{\footnotesize{School of Space Research, Kyung Hee University, Yongin, Gyeonggi 446-701, Korea}}}\\
}%
\ifthenelse{\boolean{dc}}
{
\twocolumn[
\begin{@twocolumnfalse}
\amstitle

\begin{center}
\begin{minipage}{13.0cm}
\begin{abstract}
	\myabstract
	\newline
	\begin{center}
		\rule{38mm}{0.2mm}
	\end{center}
\end{abstract}
\end{minipage}
\end{center}
\end{@twocolumnfalse}
]
}
{
\amstitle
\begin{abstract}
\myabstract
\end{abstract}
\newpage
}
\section{Introduction}
\thispagestyle{empty}
\label{sec:intro}

There is substantial evidence to suggest that changes in the solar irradiance influence variations in the temperature and circulation of the Earth's atmosphere over the 11-year solar cycle. Many of these results are based on correlations with the 10.7 cm solar flux (e.g.~\cite{LabitzkevanLoon1995,VanLoonShea1999}) or the wavelength-\\integrated, or total, solar irradiance (TSI); see \cite{Haigh2003} and references therein. While TSI is a good indicator of the total solar forcing on the climate, it cannot be used to understand the physical interaction between the solar radiation and the atmosphere since spectral solar irradiance (SSI) variability, and the altitude in the atmosphere at which it is absorbed, is highly wavelength-dependent \citep{Meier1991,LeanRottman1997,KrivovaSolanki2006}.

There is a growing body of evidence to suggest that TSI, and as a consequence SSI, may vary on secular timescales exceeding the 11-year solar cycle. \cite{Frohlich2009b}, \\ \cite{LockwoodBell2010} and \cite{BallUnruh2012} provide some evidence that TSI may have been slightly lower in the recent minimum compared to the two prior to that, though the Physikalisch- Meteorologisches Observatorium Davos (PMOD) composite of TSI observations \citep{Frohlich2006} and the modelled TSI by \cite{BallUnruh2012} are consistent, within the error bars, with no change between the last three minima. Estimates of the increase in TSI since the 17th century vary widely (see \cite{SchmidtJungclaus2012} and \cite{SolankiUnruh2013} and references therein), though most recent estimates lie in the range $\sim$1--1.5 Wm$^{-2}$ \\(\citeauthor{WangLean2005} \citeyear{WangLean2005}, \citeauthor{KrivovaBalmaceda2007} \citeyear{KrivovaBalmaceda2007}, \citeauthor{SteinhilberBeer2009} \citeyear{SteinhilberBeer2009}, \citeauthor{KrivovaVieira2010} \citeyear{KrivovaVieira2010}), a change similar to solar cycle variability.

A large proportion of the variability in TSI is due to very much larger relative variations at UV wavelengths compared to the longer visible and infra-red (IR) wavelengths. Wavelengths shorter than 400 nm account for less than 10\% of the absolute value of TSI, but contribute 30--60\% to TSI variability, according to models \citep{LeanRottman1997, KrivovaSolanki2006} and measurements by the SOLar Stellar Irradiance Comparison Experiment (SOLSTICE) \citep{RottmanWoods2001} and the Solar Ultraviolet Spectral Irradiance Monitor (SUSIM) \citep{FloydCook2003, MorrillFloyd2011} on the Upper Atmosphere Research Satellite (UARS), made prior to 2006, and the SCanning Imaging Absorption SpectroMeter for Atmospheric CHartographY \\(SCHIAMACHY) on the ENVIronmental SATellite (ENVISAT) \citep{PagaranWeber2009}. The data from the Spectral Irradiance Monitor (SIM) \citep{HarderFontenla2005a} and SOLSTICE \citep{SnowMcClintock2005} instruments on board the SOlar Radiation and Climate Experiment (SORCE) indicate that this contribution might be as high as 180\% \citep{HarderFontenla2009}, though it should be noted that SIM is currently undergoing a reanalysis. A value larger than 100\% is possible because the SSI in the visible measured by SIM varies in antiphase to that in the UV.

The UV radiation influences many processes in the atmosphere. Of particular interest is the interaction between solar UV radiation and ozone, which is the largest contributor to heating in the stratosphere. Variation of solar UV radiation over secular timescales may have an effect on global temperature trends and the impact is important to quantify. 

\cite{HaighWinning2010} and \cite{MerkelHarder2011} both investigated the potential impact that the SSI changes observed by SORCE \citep{Rottman2005, HarderFontenla2009} could have on stratospheric ozone concentrations, respectively using a coupled chemistry climate 2D atmospheric model and the fully 3D general circulation Whole Atmosphere Community Climate Model (WACCM). Both studies obtained qualitatively similar results when using hybrid SORCE data from SOLSTICE and SIM, though the two studies adopted different wavelengths to transition between SOLSTICE and SIM. While the magnitude and the exact heights varied, both studies found that between 2004 and 2007, when solar UV output was declining, ozone concentrations increased above $\sim$45 km while they decreased below $\sim$40 km. It is interesting to note that trends in O$_{3}$ from the Microwave Limb Sounder (MLS) on the Aura satellite and the Sounding of the Atmosphere using Broadband Emission Radiometry (SABER) on Thermosphere Ionosphere Mesosphere Energetics Dynamics (TIMED) observations presented by \cite{HaighWinning2010} and \cite{MerkelHarder2011}, respectively, suggest that SORCE SSI may better capture solar variability than models due to the negative response in mesospheric ozone that is out-of-phase with solar irradiance changes. On the other hand, \cite{AustinTourpali2008} did not find a negative ozone response to cycle changes of the Sun using combined data from several satellites prior to 2004. \cite{DhomseChipperfield2013} suggest that the negative response in the lower mesosphere cannot be used to distinguish between SSI datasets due to the large uncertainties in the ozone observations.

The significant differences in SSI variability between SORCE data and different models, the latter partly relying on earlier observations (see \citeauthor{ErmolliMatthes2013}, \citeyear{ErmolliMatthes2013}), indicate that there is still much uncertainty in our knowledge of how the Sun's irradiance varies spectrally. The larger UV irradiance variability, and an inverse solar-cycle trend in the visible measured by the SIM instrument on the SORCE satellite \citep{HarderFontenla2005a, HarderFontenla2009}, may indicate that the solar cycle variability observed by previous missions, and the models that reproduce similar behaviour, may be incorrect. It may also indicate a change in the Sun during the recent cycle. However, recent studies suggest that incomplete accounting for instrument degradation may contribute to the SSI trends suggested by SIM data (\citeauthor{BallUnruh2011} \citeyear{BallUnruh2011}, \citeauthor{DelandCebula2012} \citeyear{DelandCebula2012}, \citeauthor{LeanDeland2012} \citeyear{LeanDeland2012},\\ \citeauthor{ErmolliMatthes2013} \citeyear{ErmolliMatthes2013}).

This paper presents an extended and recalibrated data set using the Spectral And Total Irradiance REconstruction (SATIRE-S) model \citep{FliggeSolanki2000, KrivovaSolanki2003, WenzlerSolanki2004, BallUnruh2012}, for wavelengths between 115 and 160\,000 nm and on all days between August 1974 and October 2009, for use by the climate and atmospheric research communities. 

SATIRE-S is the most detailed of the SATIRE family of models, where `S' refers to the model designed for the satellite era \citep{KrivovaSolanki2011}, and provides the most reliable reconstruction of TSI and SSI. It does, however, rely on the availability of magnetograms and continuum intensity images, which restricts its applicability to a comparatively short period of time. In the past, the TSI and SSI reconstructed with SATIRE-S have been further limited by the fact that magnetograms obtained from different instruments do not have the same spatial resolution, noise level or magnetic field calibration. It requires careful intercalibrations between various magnetographs (and imagers) to allow a homogeneous reconstruction of TSI. This has been successfully done by \cite{WenzlerSolanki2004, WenzlerSolanki2006} for Kitt Peak Solar Observatory (KP) 512-channel magnetograph (512) \citep{LivingstonHarvey1976, LivingstonHarvey1976b} and spectromagnetograph (SPM) \citep{JonesHarvey1992} instruments and by \cite{BallUnruh2012} for 512, SPM and MDI instruments \citep{ScherrerBogart1995}, the latter of which is onboard the Solar and Heliospheric Observatory (SoHO) spacecraft.

Here we compute the SSI using magnetograms from all three instruments, thus extending the SSI reconstructed by SATIRE-S to fully cover the last three solar cycles including, for the first time, the extended solar minimum in 2008. The reconstruction is compared with the NRLSSI model \citep{Lean2000, LeanRottman2005} and data from the SOLSTICE instrument \citep{McClintockRottman2005}, onboard the SORCE satellite. We then show how the different spectral irradiances of these datasets affect changes in stratospheric O$_{3}$ using the atmospheric model based on \cite{HarwoodPyle1975}.

\section{Modelling solar irradiance with SATIRE}
\label{satmodel}
The SATIRE-S model \citep{FliggeSolanki2000, KrivovaSolanki2003, WenzlerSolanki2006, KrivovaSolanki2011} assumes that all irradiance variations are the result of changes in the surface photospheric magnetic flux. SATIRE-S identifies four solar surface components in magnetograms and continuum intensity images: the background quiet sun; the dark penumbral and umbral components of sunspots; and small-scale magnetic features, that appear predominantly bright, called faculae.

Daily irradiance spectra are produced by summing the intensities of the four components weighted according to their surface distribution. The component intensities (as functions of wavelength and limb angle) are calculated with the spectral synthesis program ATLAS9 \citep{Kucurz1993} assuming local thermodynamic equilibrium (LTE) conditions. We use time-independent model atmospheres  \\ \citep{FliggeSolanki2000, KrivovaSolanki2003, SolankiUnruh2013} with effective temperatures of 5777~K, 5450~K and 4500~K for quiet Sun, penumbral and umbral intensities, respectively. For faculae, we use the FAL-P model atmosphere \citep{FontenlaAvrett1993}, as modified by \cite{UnruhSolanki1999}. The wavelength grid of the daily spectra from SATIRE-S has a resolution of 1 nm below 290 nm, 2 nm from 290 to 1000 nm, 5 nm from 1000 to 1600 nm, 10 nm from 1600 to 3200 nm, 20 nm from 3200 to 6400 nm, 40 nm from 6400 to 10\,000 nm and 20\,000 nm for the remainder of the spectrum up to 160 $\mu$m.

The model has one free parameter; this relates the magnetic flux registered in a magnetogram pixel to the fraction of the pixel filled by faculae. The free parameter is set to a fixed value for each observatory (i.e., for KP and SoHO) as outlined in the next section.

\subsection{Method to combine reconstructions}
\label{combine}

In order to maximise the length of the SSI timeseries, magnetograms and continuum intensity images are taken from three instruments: two at KP that are based on spectropolarimetry of the Fe~{\sc i} 868.8~nm line \citep{LivingstonHarvey1976}, the KP/512 \citep{LivingstonHarvey1976b} and KP/SPM instrument \citep{JonesHarvey1992}, and the SoHO/MDI instrument that uses the Ni~{\sc i} 676.8~nm line \citep{ScherrerBogart1995}. The free parameter for each instrument is fixed by comparing the reconstructed TSI to either TSI observations or to a TSI reconstruction made using images from a different instrument. Broadly, three steps are involved in the inter-calibration, which are outlined below (see also \citeauthor{BallUnruh2012}, \citeyear{BallUnruh2012}). The uncertainties arising in this process are outlined in section~\ref{satmodel}\ref{uncert}.

In step (i), we fix the free parameter for the MDI reconstruction by requiring a regression slope of unity between the reconstructed TSI and the SORCE Total Irradiance Monitor (TIM) TSI observations \citep{KoppLawrence2005}. In step (ii), we combine the KP and MDI magnetogram and continuum images using the KP/SPM and SoHO/MDI overlap period of 895 days between 1999 and 2003. This requires fixing the free parameter for SPM, so that the reconstructed TSI during the overlap period agrees with the TSI derived from the MDI images. However, while the resulting spectral irradiances are very well correlated for the overlap period ($r_c > 0.91$ at all wavelengths), we see slightly different variability amplitudes in the two reconstructions at some wavelengths. The different instrument design, the use of different spectral lines with different magnetic sensitivities, as well as different telescope optics and detectors mean that there are non-linear, position-dependent differences in the KP and MDI instruments in response to magnetic flux that leads to the differing variability amplitude. The differences in amplitude of variability in the overlap period are typically 2\% in the visible and near-IR and remain below 8\% at all wavelengths. To avoid discontinuites in the SSI trends when changing between reconstructions based on MDI and KP magnetograms, we adjust the variability amplitudes of the KP reconstructions by rescaling them to those of the MDI reconstructions.

Step (iii) involves correcting for the change between the 512 and SPM instruments on KP. While the imaging quality for the KP/512 data is poorer, the two KP polarimeters show very similar flux registration so that the correction can be used to convert the KP/512 magnetogram signal to the KP/SPM level (see \citeauthor{WenzlerSolanki2006}, \citeyear{WenzlerSolanki2006}, and \citeauthor{BallUnruh2012}, \citeyear{BallUnruh2012}). Thus, the same filling factor can be used for both KP data sets. While the scaling factor introduces uncertainties regarding the long-term TSI behaviour, it does not affect its spectral distribution.

The ATLAS9 model intensities assume LTE conditions in the solar atmosphere; this can result in large errors in the modelled irradiance variability in some wavelength regions, mainly below 270 nm and at the Mg I line at 285 nm. SATIRE-S does show, however, good agreement with SSI observations from the UARS satellite: the reconstructed SSI in the range 220--240 nm agrees well with UARS/SUSIM measurements \citep{KrivovaSolanki2006, KrivovaSolanki2009} and reasonably well with UARS/SOLSTICE measurements \citep{UnruhBall2012}.

To better reflect the spectral irradiance variability between 115 nm and 270 nm, we apply the empirical method outlined in \cite{KrivovaSolanki2006}. This method relies on the good agreement in the temporal variability of the 220 to 240 nm region as calculated by SATIRE-S and uses the scaling coefficients derived from spectral irradiance measurements, over the period 1997 to 2002, taken by the UARS/SUSIM instrument (\citeauthor{BruecknerEdlow1993} \citeyear{BruecknerEdlow1993}, \\ \citeauthor{FloydCook2003} \citeyear{FloydCook2003}). Therefore, spectral regions in SATIRE-S below 220 nm and between 240 and 270 nm rely on SUSIM measurements and the close agreement in these regions is partly by design. In section~\ref{timeseries} we show an example of this with the reconstructed Ly-$\alpha$ irradiance, which is in agreement with the composite of Ly-$\alpha$ measurements and proxy-models by \cite{WoodsTobiska2000}, and three integrated UV wavelength bands below 290 nm.

The TSI data set that is obtained by integrating our new SSI reconstruction is considered as an update of the TSI reconstruction presented in \cite{BallUnruh2012}. While both reconstructions are consistent within their uncertainty ranges, the updated TSI is now based upon the integral of the SSI that is self-consistent at every wavelength for the full reconstruction period. In \cite{BallUnruh2012}, the inter-cycle decline between 1996 and 2008 was estimated to be 0.20$^{+0.12}_{-0.09}$ Wm$^{-2}$, where the errors are one-sigma uncertainties. The new reconstruction revises this estimate down to 0.13$^{+0.07}_{-0.10}$ Wm$^{-2}$. Note that the reconstructions are calibrated using only the SORCE/TIM measurements and are thus independent of any TSI composite post-1990 and independent within the uncertainty range prior to this period (see \citeauthor{BallUnruh2012}, \citeyear{BallUnruh2012}, for more details).

In a final step we adjust the absolute levels of SATIRE-S SSI so that the integrated SSI is in agreement with SORCE/\\TIM at the solar minimum in December 2008. For this, the entire spectrum is multiplied by a factor of 1.0047. This small correction of 0.5\% assures that the original variability as obtained directly from SATIRE-S, is not affected.

\subsection{Data gap filling}
\label{gapfill}
The new SATIRE-S SSI reconstruction now extends through the most recent and unusually long solar minimum period, whereas the previous version \citep{KrivovaSolanki2009} ended in 2007. For the period between 1974 December 10 and 2009 October 31, images are missing on $\sim$50\% of dates, mostly within cycles 21 and 22; we fill these data gaps to provide fluxes on all dates over the entire period. To avoid any assumptions about the solar behaviour, we decompose each wavelength into short-term, or rotational, and long-term timeseries. The long-term timeseries is obtained by smoothing the original timeseries using a Gaussian window equivalent to a boxcar width of 135 days. We use this period, longer than the typical 81-days, to reduce the impact of short-term variability. The short-term timeseries, which captures rotational variability, is obtained by subtracting the long-term timeseries from the original.

Gaps in the long-term SATIRE-S time-series are filled by linear interpolation. Most gaps are short, with 90\% of gaps being no longer than a solar rotation of 27 days in length, so the long-term trend is well approximated by a linear interpolation. Only five data gaps exceed two solar rotations, the longest of which is a 282 day period around the solar minimum of 1976.

Gaps in the detrended, rotational time series are filled using solar activity indices: the NOAA and LASP Mg II indices \citep{ViereckFloyd2004,SnowMcClintock2005}, combined through linear regression; the Ly-$\mathrm{\alpha}$ composite by \cite{WoodsTobiska2000}; the Penticton F10.7 cm radio flux (data available through the National Geophysical Data Center at \\ http://www.ngdc.noaa.gov/); the TSI from version \\ d41\_62\_1003 of the PMOD composite \citep{Frohlich2000} and the sunspot area composite record by \cite{BalmacedaSolanki2009}. Each index is indicative of the behaviour of some feature in the solar atmosphere, although it is not clear exactly how they relate to each wavelength of SSI (see \citeauthor{DudokdeWitKretzschmar2008}, \citeyear{DudokdeWitKretzschmar2008}, and supplementary material section~b). Rotational variability at each wavelength is better approximated using multi-linear regression of two indices than using just one. We calculated the regression coefficient for every combination of two indices for each wavelength using dates when all indices and reconstructed SSI exist. Then, for each data gap, the available index-pair with the highest value of the coefficient of determination, r$_{\mathrm{c}}^{\mathrm{2}}$, at each wavelength is used to calculate the SSI in the missing gap. 

Finally, the detrended and smoothed time series are added together to produce a spectral reconstruction that reflects the long-term variability of SATIRE-S while retaining rotational consistency (see supplementary material section~b for examples). This procedure is expected to perform less well prior to 1978 because the TSI and the Mg II index, which generally have the highest combined correlation coefficients, are unavailable then. Also, this time period coincides with the longest data gaps in the reconstruction.

\begin{figure*}
\begin{center}
 \noindent\includegraphics[width=43pc]{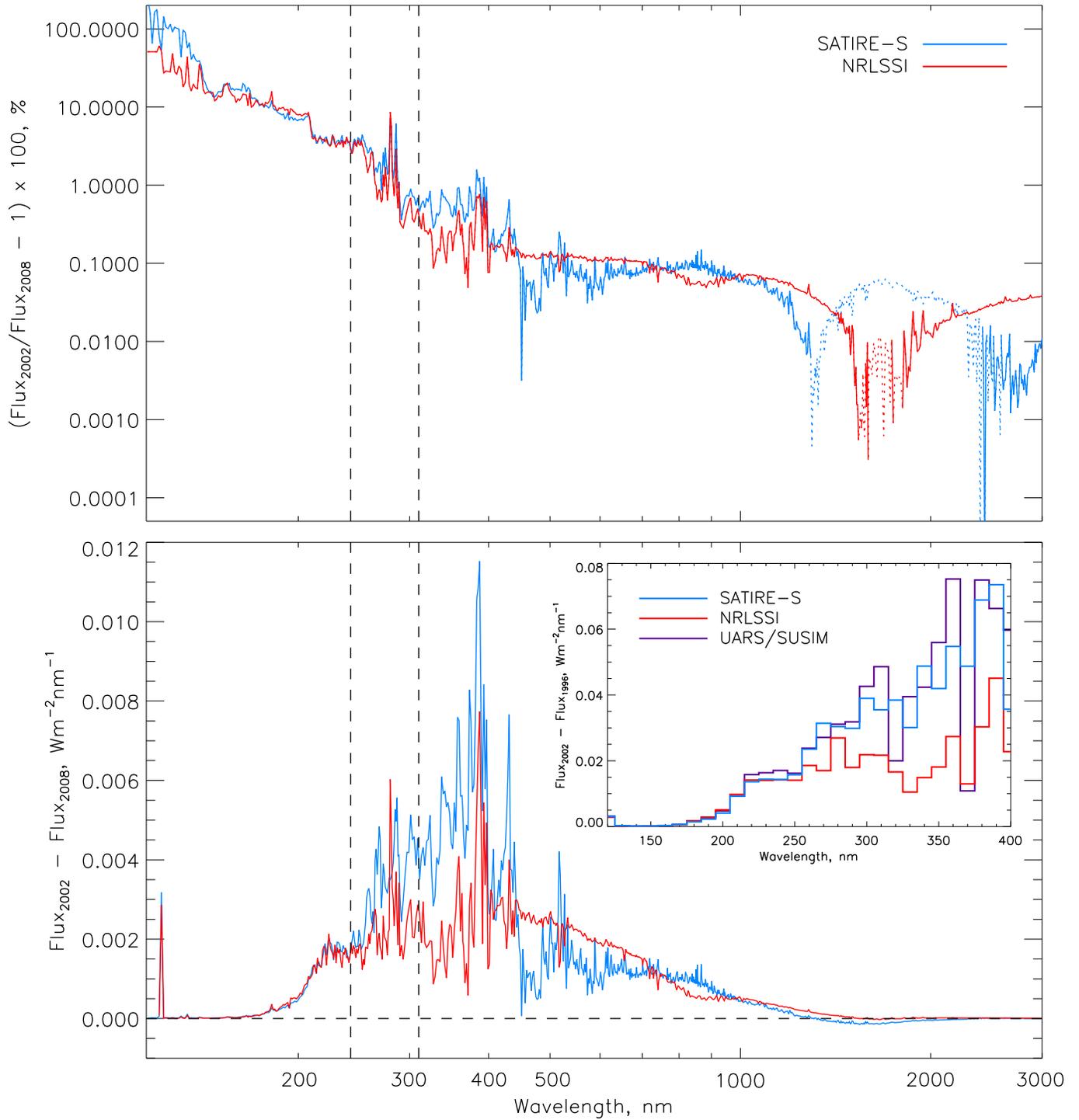}
 \caption{The (top) percentage and (bottom) absolute change in flux between cycle 23 maximum and minimum, in terms of the maximum change in irradiance between three-month averages at February 2002 and December 2008 (see Fig.~\ref{tsi}), for the SATIRE-S (blue) and NRLSSI (red) models; in the upper plot solid lines represent a decrease in flux while dotted lines represent an increase. The vertical dashed lines indicate 242 and 310 nm and the horizontal line marks no change in flux. (inset) The absolute change in flux in 10 nm bands between three month averages at May 1996 and February 2002 for SATIRE-S (blue), NRLSSI (red) and UARS/SUSIM (purple).}
 \label{fig7}
\end{center}
\end{figure*}

The change in SSI between maximum activity in solar cycle 23, in 2002, and the minimum in 2008 is plotted in Fig.~\ref{fig7}. The blue curve indicates the final SATIRE-S reconstruction. This figure is described in greater detail and discussed in section~\ref{compare}\ref{compnrl}.

\subsection{Uncertainty estimate}
\label{uncert}

An accurate error estimate for the modelled reconstruction is difficult to provide, since it depends partly on unknowns (such as the amount of magnetic flux missed by the magnetograms) and on uncertainties that cannot be precisely constrained within the scope of this paper (such as the accuracy of the model atmospheres employed, or the influence of neglecting non-LTE effects; see section~c of the supplementary materials). Therefore, we attempt to provide a long-term SSI uncertainty range similar to the approach taken for TSI in \cite{BallUnruh2012}. This is an empirical approach that takes into account the uncertainties introduced in the calibration steps described in section~\ref{satmodel}\ref{combine}. Specifically, we account for step (i), the regression fitting between the TSI derived from MDI images and the SORCE/TIM measurements, step (ii), the regression fitting to combine the SATIRE-S reconstructions for MDI and KP data, and step (iii), the uncertainties in the correction factor for the KP/512 relative to the KP/SPM magnetograms.

For wavelengths below 270 nm we add, in quadrature, the uncertainty from the SATIRE-S reconstruction and the estimated relative uncertainty of the UARS/SUSIM measurements, which are estimated to be of the order of 5\% below 142 nm and decreases to about 2\% above 160 nm (\citeauthor{WoodsPrinz1996}, \citeyear{WoodsPrinz1996}, and Linton Floyd, personal communication); we linearly interpolate the uncertainty between these wavelengths. These uncertainties are provided with the published reconstruction. We also flag a few wavelengths in the SATIRE-S spectrum, most notably the Mg I line at 285 nm, where our detailed comparisons with SORCE/SIM measurements on rotational time scales indicate that SATIRE-S overestimates the solar variability; see section~c of the supplementary materials.

To summarise and illustrate the temporal behaviour of the uncertainties, we list the cycle amplitudes and their associated uncertainties in Tab.~\ref{tab1} for the TSI and for selected `broadband' spectral irradiances. Column 4 lists the cycle amplitudes for cycle 23; these are based on reconstructions from SoHO/MDI images, i.e. these account for uncertainties in step (i) only. As shown in \cite{BallUnruh2012}, the error on the regression fitting for the SoHO/MDI free parameter is small and arises mainly from the long-term uncertainty in SORCE/\\TIM. The uncertainty on the cycle amplitude (between sunspot maximum in March 2000 to the minimum in December 2008), is of the order of 100 ppm, comparable to the long-term uncertainty of SORCE/TIM. We note that the agreement between the TSI derived with SATIRE-S and the available composites is just as good as the agreement between the different composites \citep{BallUnruh2012}.

\begin{figure*}[t]
\begin{center}
 \noindent\includegraphics[width=17pc, angle=90]{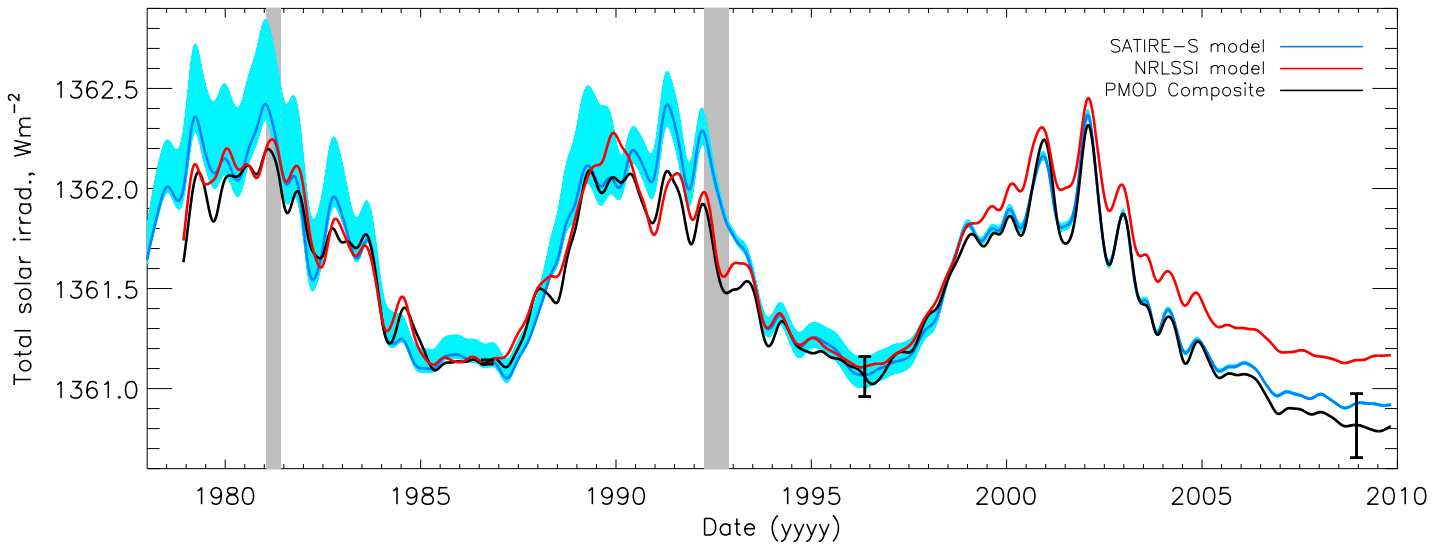}
 \caption{Smoothed (solid) wavelength-integrated time series of SATIRE-S (blue), NRLSSI (red) and the PMOD composite of TSI (black) between 1978 and 2010. PMOD and NRLSSI have been normalised to the minimum of SATIRE-S in 1986. The SATIRE-S uncertainty range is shown by light blue shading and the PMOD error bars are given at the cycle minima. Grey shading highlights the periods where gaps exceeding three solar rotations have been filled.}
 \label{tsi}
\end{center}
\end{figure*}

Going back in time, the uncertainties increase, mainly due to the additional calibration steps (ii) and (iii). This is illustrated by the larger uncertainties for the amplitudes of cycles 21 and 22 (see columns 2 and 3 of Tab.~\ref{tab1}, respectively). As illustrated in Fig.~\ref{tsi} and Fig.~7 in \cite{BallUnruh2012}, the uncertainties are asymmetric and typically show slightly larger positive ranges. This is due to the different response of the magnetograms when connecting reconstructions from MDI and KP, as in step (ii), and the uncertainties in the correction factor from step (iii) that only lead to an increase in flux variability, not a decrease (see \citeauthor{BallUnruh2012}, \citeyear{BallUnruh2012}).

\begin{table*}[t]
\label{tab1}
\caption{Amplitude of flux variability over a solar cycle (SC) for selected wavelength bands and for total solar irradiance (TSI). The uncertainty of the amplitude is also given.}
\begin{center}
\begin{tabular}{llll}
\hline
$\lambda\lambda$, nm & SC21, mWm$^{-2}$ & SC22, mWm$^{-2}$ & SC23, mWm$^{-2}$ \\
\hline
200--270 & 114$^{+25}_{-8}$ & 114$^{+25}_{-8}$ & 99$^{+4}_{-4}$  \\
270--400 & 509$^{+127}_{-46}$ & 501$^{+135}_{-47}$ & 445$^{+16}_{-21}$  \\
400--700 & 241$^{+141}_{-72}$ & 190$^{+166}_{-80}$ & 201$^{+37}_{-51}$  \\
700--1000& 138$^{+67}_{-34}$ & 116$^{+78}_{-38}$ & 118$^{+18}_{-24}$  \\
TSI      & 979$^{+394}_{-183}$ & 876$^{+445}_{-195}$ & 837$^{+85}_{-113}$  \\
\hline
\end{tabular}
\end{center}
\end{table*}

While considerable progress has been made in determining the absolute value of the total solar irradiance \\ \citep{KoppLean2011}, the absolute spectral solar irradiance is still poorly constrained and a number of different `standard' absolute solar spectra are available (see \cite{ThuillierHerse2003} for a discussion of this). For this reason, the uncertainties listed here and distributed with the reconstructions are for relative irradiances. Relative spectral irradiances in each band are much better constrained than the absolute accuracy, though the degradation of space instruments means that considerable uncertainties remain when going beyond rotational time scales \citep{UnruhBall2012,ErmolliMatthes2013}. If users of the SATIRE-S reconstruction wish to use a different absolute spectra as a basis upon which SATIRE-S variability is placed, we provide absolute spectra binned onto the SATIRE-S wavelength grid. We do this for the ATLAS 3 and Whole Heliosphere Interval solar reference spectra \citep{WoodsChamberlin2009}.

The SATIRE-S SSI data can be found at http://www. mps.mpg.de/projects/sun-climate/data.html.

\section{Inter-comparison of SATIRE-S with other \\datasets}
\label{compare}

In this section, we compare the SATIRE-S model with the NRLSSI model \citep{Lean2000, LeanRottman2005} and with SORCE/SOLSTICE observations. NRLSSI is the most widely used, empirically derived, model of SSI, so we perform a comparison with SATIRE-S here. The solar cycle spectral variability recorded by SORCE/SOLSTICE is larger than the variability seen by the UARS/SUSIM and UARS/SOLSTICE instruments. It is also larger than the variability inferred from the models. SORCE/SIM displays even larger variability than SORCE/SOLSTICE, for the overlapping range between 240--310 nm, and it is an interesting SSI dataset to consider. However, a comparison of SATIRE-S and an updated version of SORCE/SIM version 17 was made in \cite{BallUnruh2011}. The SORCE/SIM UV changes below 310 nm were up to five times larger than the changes seen in SATIRE-S. We note that Figure 11 of \cite{BallUnruh2011} indicates that SORCE/SIM displays up to 10 times the variability of SATIRE-S at some wavelengths between 300 and 400 nm, though the integrated variability of SORCE/SIM over this wavelength range was approximately 3.4 times larger than SATIRE-S. \cite{BallUnruh2011} suggested that degradation of the instrument has not been properly accounted for and may be overestimating the long-term trends (see also \cite{LeanDeland2012, DelandCebula2012, UnruhBall2012, ErmolliMatthes2013}). SORCE/SIM is currently undergoing a reanalysis that may affect the cycle variability and its uncertainty estimates (Peter Pilewskie, personal communication). We, therefore, only consider SORCE/SOLSTICE in the following. We compare SATIRE-S and NRLSSI for wavelengths between 120 and 3000 nm and up to 310 nm when comparing with SORCE/SOLSTICE.

\subsection{Comparison with the NRLSSI model}
\label{compnrl}

NRLSSI is an empirical model that uses the disk-\\integrated Mg II and photospheric sunspot indices to describe the evolution of sunspots and faculae, respectively. For wavelengths $<$ 400 nm, spectral irradiances are computed from multiple regression analysis with UARS/\\SOLSTICE observations. This analysis is performed on detrended, rotational data to avoid instrumental degradation effects and, therefore, assumes that rotational variability scales with solar cycle changes in irradiance. For wavelengths above 400 nm, facular and sunspot contrasts, from the models by \cite{SolankiUnruh1998}, are scaled to agree with solar cycle TSI observations. NRLSSI's integrated flux is $\sim$4 Wm$^{-2}$ higher than SORCE/TIM so we normalise NRLSSI to SORCE/TIM, as was done for SATIRE-S in section~\ref{satmodel}\ref{combine}.

It is worth briefly considering how the TSI derived by integrating the SSI differs between the two models. In Fig.~\ref{tsi}, the smoothed, wavelength-integrated SATIRE-S \\(blue), NRLSSI (red) and PMOD composite of TSI (black) \citep{Frohlich2003} are plotted between 1978 and 2009. NRLSSI and PMOD have been normalised to the absolute value of SATIRE-S averaged over three months centred at the minimum of 1986. The SATIRE-S uncertainty range is plotted with light blue shading and the cycle minima error bars from PMOD are plotted as black bars \citep{Frohlich2009b}. Although other TSI composites exist, PMOD is now generally accepted as the most accurate TSI composite of observations, which is why we consider only it here; see \cite{BallUnruh2012} for comparisons with all TSI composites on rotational and cyclical timescales. We find correlation coefficients between NRLSSI and SATIRE-S TSI with PMOD, of 0.92 and 0.96, respectively; detrending the timeseries, as described in section~\ref{satmodel}\ref{gapfill}, yields correlation coefficients of 0.87 and 0.96 for the rotational variability. These statistics suggests that SATIRE-S reproduces the PMOD composite of observations better than NRLSSI. However, there are periods when NRLSSI matches PMOD better, on yearly and longer timescales, than SATIRE-S. For example, the large difference between SATIRE-S and PMOD around 1991-1993 is due to the remaining uncertainties in the cross-calibration of the KP 512 and SPM magnetograms (see also \citeauthor{WenzlerSolanki2006}, \citeyear{WenzlerSolanki2006}, and \citeauthor{BallUnruh2012}, \citeyear{BallUnruh2012}).

The inter-cycle trends of the three datasets are subtly different. Whereas PMOD and SATIRE-S show a decline of $\sim$0.20$^{+0.16}_{-0.26}$ and 0.13$^{+0.07}_{-0.10}$ Wm$^{-2}$, respectively, between 1996 and 2008, NRLSSI exhibits no change over this period. NRLSSI's behaviour is most likely due to the use of the Mg II index as a proxy for long-term changes; this index does not exhibit any strong inter-cycle variation \citep{Frohlich2009b}. Note, however, that the uncertainty of the cycle minima in PMOD and SATIRE-S also encompasses the NRLSSI model estimate of no change, though the Mg II record is not entirely free of long-term uncertainty either (\citeauthor{SnowMcClintock2005},~\citeyear{SnowMcClintock2005}; Marty Snow, personal communication). We note that while accurate TSI is important to act as a constraint for the SSI in both NRLSSI and SATIRE-S (and SORCE/SIM, which is much less well constrained, see \cite{BallUnruh2011}), it does not ensure that the SSI is correct, in either case, as higher variability at some wavelengths can be compensated by lower variability at other wavelengths.

For the SSI comparison between NRLSSI and SATIRE-S, in Fig.~\ref{fig7} we consider the change in flux, $\Delta$F, between two 81-day averaged periods centred on: 2002 February 1, the second and highest peak of cycle 23; and the cycle 23/24 minimum on 2008 December 15. This provides the largest range of change in cycle 23.

In the upper plot of Fig.~\ref{fig7} the percentage change between the maximum and minimum of solar cycle 23 is plotted on a logarithmic scale while the lower plot depicts the absolute change in flux on a linear y-axis. The spectral uncertainty in SATIRE-S is very small for cycle 23 and is virtually invisible on the plotted scales. We, therefore, plot the spectral variability and the uncertainties for each cycle in section~a of the supplementary materials. The wavelength spacings in the two models are different. Consequently, NRLSSI has been interpolated onto the SATIRE-S wavelength grid (see section~\ref{satmodel}). The regions below 242 nm and between 242 and 310 nm are important in ozone production and destruction processes in the stratosphere (see section~\ref{2dmodel}), so these have been highlighted with vertical dashed lines.

Below 242 nm, the two models agree well in the change of flux, with larger differences apparent only below $\sim$150 nm; these result from the use of different instruments to define the cycle variability, i.e. UARS/SUSIM for SATIRE-S (see section~\ref{satmodel}\ref{uncert}) and the scaled rotational variability of UARS/SOLSTICE for NRLSSI (see above). Integrating over 120--242 nm, a region important for the photodissociation of O$_{2}$ and O$_{3}$, we find that NRLSSI shows the same solar cycle change as SATIRE-S, though we note that the SATIRE-S Lyman-$\alpha$ has $\sim$6\% larger cycle amplitude than NRLSSI during the descending phase of cycle 23, but which is generally larger for NRLSSI in earlier solar cycles (see section~\ref{timeseries} and Fig.~\ref{figslyman}). These differences in Lyman-$\alpha$ cycle changes will impact on OH chemistry that indirectly affect ozone concentrations. Between 242 and 310 nm, important in O$_3$ photodissociation, the integrated change in flux, $\Delta$F, is more than 50\% larger in SATIRE-S than NRLSSI.

SATIRE-S shows up to three times larger cycle variability than NRLSSI for wavelengths between 300 and 400 nm. At longer wavelengths, between 400 and 1250 nm, i.e. in the visible and near-IR, NRLSSI generally displays larger cycle variability than SATIRE-S. At yet longer wavelengths, both NRLSSI and SATIRE-S display negative variability in the IR, though this occurs over a wider range and with higher variability in SATIRE-S than NRLSSI, the latter of which only shows negative variability between 1500 and 1850 nm. Where this transition to negative variability occurs is highly dependent on the facular model used (see supplementary materials section~c and \cite{UnruhSolanki1999, UnruhSolanki2000,UnruhKrivova2008}). We note that negative variability in the IR region is also seen in SORCE/SIM in the IR above 970 nm, but this change is much larger than both NRLSSI and SATIRE-S. In \cite{BallUnruh2011}, the integrated IR region of 972-1630 nm shows $\sim$0.00 Wm$^{-2}$ change in SATIRE-S between 2004 and 2008, while SORCE/SIM increases by 0.24 Wm$^{-2}$. For all wavelengths between 200 and 1600 nm, detailed comparisons between SORCE/SIM and the SATIRE-S and NRLSSI models are made by \cite{BallUnruh2011} and \cite{LeanDeland2012}, respectively.

\subsection{Comparison with SORCE/SOLSTICE}

\begin{figure*}[t]
 \noindent\includegraphics[width=43pc, angle=0]{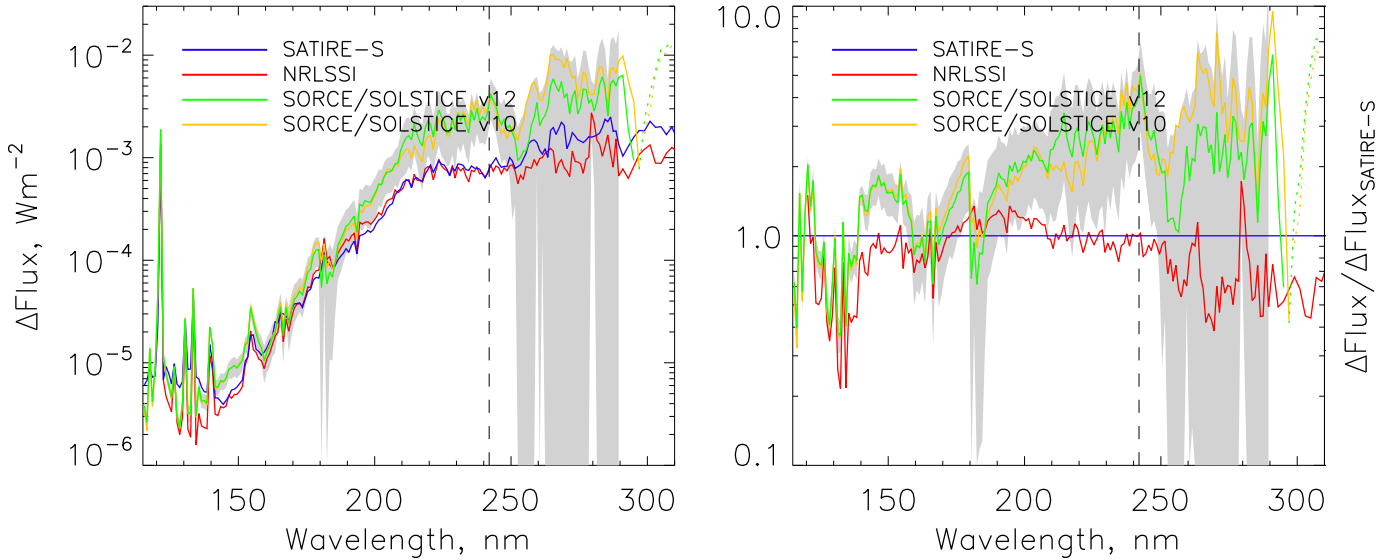}
 \caption{The SSI change between two 81-day periods centred on 2003 August 15 and 2008 December 15 (left) in absolute flux and (right) relative to SATIRE-S, on a logarithmic scale. Plotted are SORCE/SOLSTICE version 10 (orange) and 12 (green), SATIRE-S (blue) and NRLSSI (red). Dotted lines indicate an inverse trend in flux between 2003 and 2008; flux variations at wavelengths longer than 290 nm cannot be considered reliable as the long-term uncertainty exceeds the flux change, $\Delta$F, at these wavelengths (see main text). The uncertainty range for SORCE/SOLSTICE version 12 is shown in grey shading for wavelengths below 290 nm. Wavelengths are 1 nm width bins below 290 nm and 2 nm bins above. Dashed lines indicate the wavelength 242 nm.}
 \label{fig8}
\end{figure*}

SORCE/SOLSTICE observations cover the UV region 115--310 nm. In Fig.~\ref{fig8}, we compare the modelled SSI with SORCE/SOLSTICE between two 81-day average periods centred on 2003 August 15 and 2008 December 2008. This period constitutes $\sim$60\% of the full cycle variation in TSI and allows for a direct comparison with SORCE/SOLSTICE observations that started on 2003 May 14, i.e. some time after the maximum of cycle 23. The left plot of Fig.~\ref{fig8} shows the absolute change in flux, while the right is the change in absolute flux relative to SATIRE-S. We plot version 10 of SORCE/SOLSTICE data in addition to the latest version 12 as we consider version 10 in the next section. SORCE/SOLSTICE data have a reported long-term uncertainty of 0.5\% per year which amounts to $\sim$2.7\% for the period considered in Fig.~\ref{fig8} and is shown with grey shading for version 12 only; version 12 data have additional degradation corrections to prior versions that result in a change to the absolute level and relative change in spectral irradiance (Marty Snow, personal communication). When not otherwise specified, version 12 is referred to in the following.

From Fig.~\ref{fig8}, SORCE/SOLSTICE cycle variability is in reasonable agreement with the models below 140 nm and near 165, 180 and 255 nm. At most other wavelengths below 290 nm, cycle variability is larger in SORCE/SOLSTICE than in the models, typically by at least a factor of two: it is twice that of SATIRE-S at $\sim$170 and 200 nm; between 200 and 250 nm SORCE/SOLSTICE increases from two up to five times the change in SATIRE-S; and between 260 and 290 nm it typically ranges between 1.7 and 3.5 times larger than SATIRE-S. Above 290 nm the long-term uncertainty of SORCE/SOLSTICE exceeds the flux change, $\Delta$F (Marty Snow, personal communication), so results cannot be considered realistic or useful in comparison to the models.

\subsection{Timeseries comparison with NRLSSI and \\SORCE/SOLSTICE}
\label{timeseries}

We show examples of timeseries for three different wavelength bands: Lyman-$\alpha$ at $\sim$121 nm, 176--242 nm and 242--290 nm.

\begin{figure*}[t]
 \noindent\includegraphics[width=43pc]{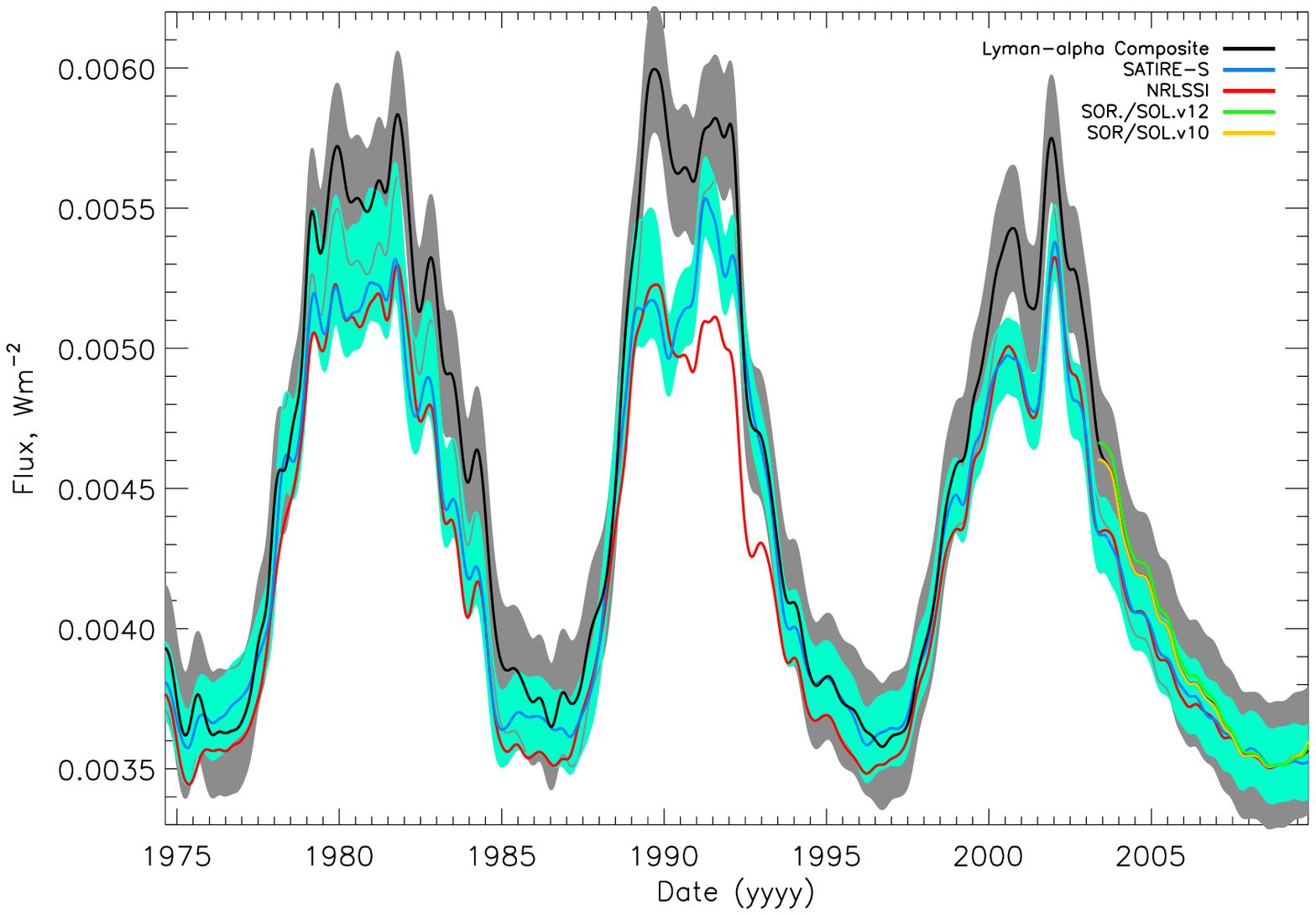}
\caption{The Lyman-$\alpha$ composite (black) by \cite{WoodsTobiska2000} with an uncertainty of 10\% of the cycle amplitude (grey shading) compared with SATIRE-S (blue) with the uncertainty range (light blue shading), the NRLSSI model (red) and SORCE/SOLSTICE version 12 (green) and 10 (yellow). SATIRE-S and NRLSSI are normalised to the composite at the solar minimum of 1986 and SORCE/SOLSTICE datasets are normalised to the composite at the solar minimum of December 2008.}
\label{figslyman}
\end{figure*}

In Fig.~\ref{figslyman}, we compare the Lyman-$\alpha$ composite by \\ \cite{WoodsTobiska2000} (black) with SATIRE-S (blue), NRLSSI (red) and SORCE/SOLSTICE version 12 (green) and 10 (yellow, dashed). In this figure, SATIRE-S and NRLSSI are normalised to the Lyman-$\alpha$ composite for the solar minimum of 1986 (as was done for the TSI plot with PMOD in Fig.~\ref{tsi}), while both versions of SORCE/SOLSTICE are normalised to the three month averaged period around December 2008. The grey shading is an estimated 10\% uncertainty \citep{WoodsTobiska2000} on the cycle amplitude, or approximately 0.22 mWm$^{-2}$ at the 1$\sigma$ level. The Lyman-$\alpha$ composite agrees with SORCE/SOLSTICE v10 almost exactly during the declining phase of cycle 23 because that version was used in the Lyman-$\alpha$ composite. Light blue shading is the uncertainty range of SATIRE-S at 121.5 nm and is the quadrature sum of uncertainty from the SATIRE-S reconstruction and an estimated 5\% uncertainty from UARS/SUSIM for the correction applied using the method by \cite{KrivovaSolanki2006}.

The nominal values of NRLSSI and SATIRE-S are at the lower end of the Lyman-$\alpha$ uncertainty range. Considering that there is uncertainty at the cycle minima and maxima, SATIRE-S and NRLSSI absolute values could be shifted up and remain within the Lyman-$\alpha$ composite uncertainty in all three solar cycles. Therefore, both models agree with the Lyman-$\alpha$ observations, within the uncertainty, at almost all times on annual to decadal time scales.

\begin{figure*}[t]
 \noindent\includegraphics[width=43pc]{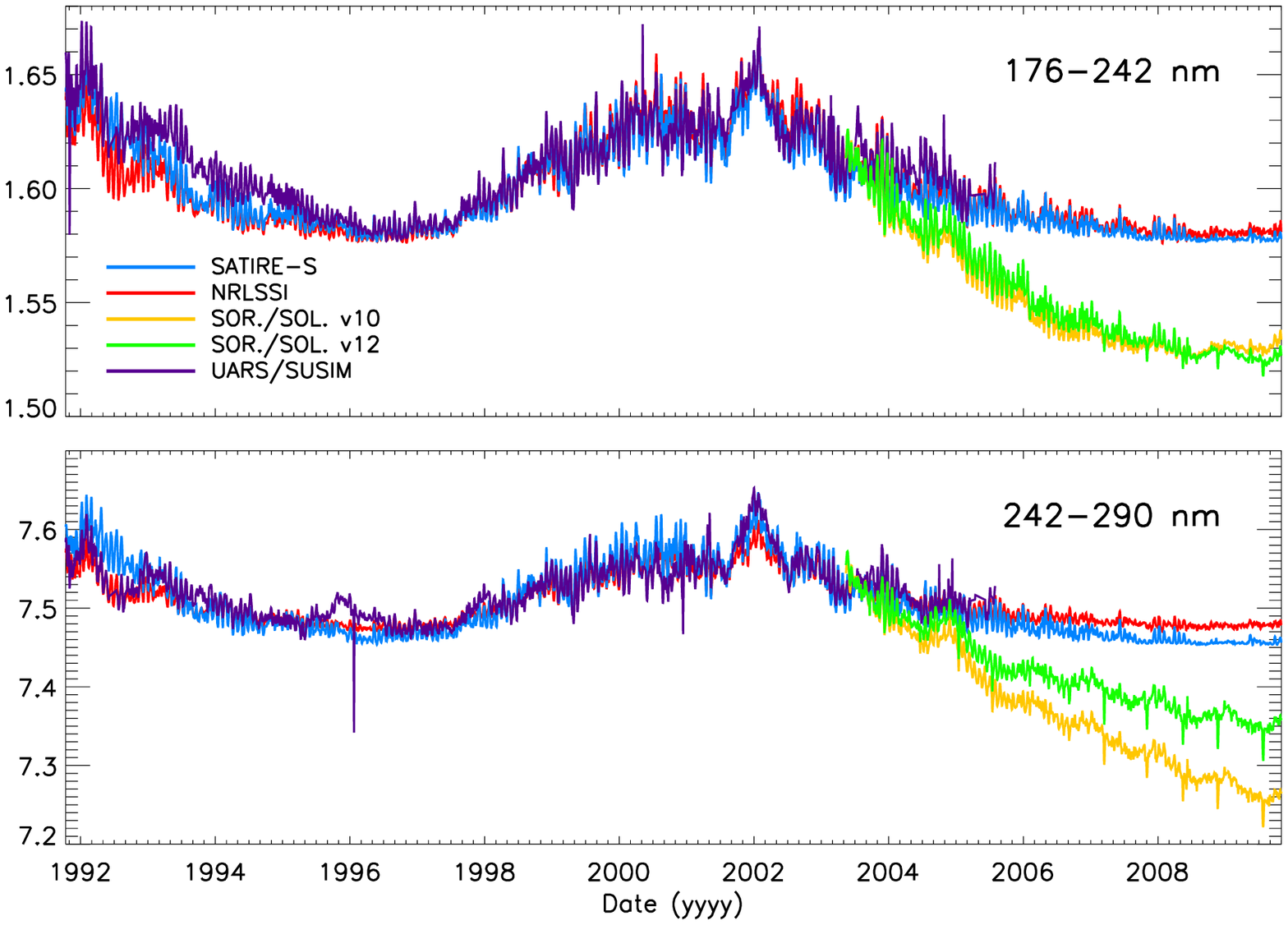}
\caption{Timeseries of SATIRE-S (blue), NRLSSI (red), SORCE/SOLSTICE version 10 (yellow) and version 12 (green) and UARS/SUSIM (purple) are shown for the wavelength bands (top) 176--242 nm and (bottom) 242--290 nm between 1991 and 2009. UARS/SUSIM and NRLSSI are shifted to the mean of SATIRE-S for the period between 1997 and 2000 and SORCE/SOLSTICE v10 and v12 between June and November 2003.}
\label{figstimeseries}
\end{figure*}

UARS/SUSIM monitored spectral irradiance between 115 and 410 nm for the period 1991 to 2005 and suggested lower solar cycle changes in UV SSI than SORCE/\\SOLSTICE. In Fig.~\ref{figstimeseries}, we show timeseries for two integrated regions, between 176--242 nm and 242--290 nm (important in the production and destruction of stratospheric ozone, respectively), for SATIRE-S (blue), NRLSSI (red), SORCE/SOLSTICE version 10 (yellow) and version 12 (green) and also UARS/SUSIM (purple). The absolute fluxes of NRLSSI and UARS/SUSIM are shifted to SATIRE-S for the average of the period between 1997 and 2000 and SORCE/SOLSTICE are shifted to SATIRE-S by the average over the period between June and November 2003. Both wavelength bands show similar solar cycle behaviour for UARS/SUSIM and the two models; SORCE/SOLSTICE show much larger solar cycle trends. We note that the degradation correction issue discussed by \cite{KrivovaSolanki2006} can be seen just prior to the solar minimum of 1996. This would lead to a change of the cycle maximum in 1991 relative to that of 2000.

However, the two bands in Fig.~\ref{figstimeseries} are broad spectral bands: selecting different wavelength ranges might lead to different conclusions as to which model is better at reconstructing SSI solar cycle changes based on UARS/SUSIM measurements. The inset in the lower plot of Fig.~\ref{fig7} shows the solar cycle change in flux from three month averages, centred on the solar minimum in 1996 and solar maximum in 2002, in 10 nm bands between 120 and 400 nm. This plot illustrates the wavelength dependence of the solar cycle changes. All three datasets show good agreement below 250 nm. Above 250 nm, UARS/SUSIM and SATIRE-S typically show larger solar cycle changes than NRLSSI; SATIRE-S shows closer solar cycle changes to UARS/SUSIM than NRLSSI.

Understanding the effect that the Sun has on Earth's atmosphere and climate requires reliable estimates of SSI variability. The differences between SATIRE-S, NRLSSI and SORCE/SOLSTICE lead to different predicted impacts on the Earth's atmosphere, e.g. heating and photodissociation rates, as the following example considering changes in ozone concentration shows.

\section{Modelled $\Delta$O$_{3}$ from different input SSI}
\label{2dmodel}

We employ the radiative-chemical transport atmospheric model based on that of \cite{HarwoodPyle1975} to compare the resultant change in ozone concentration, $\Delta$O$_{3}$, when using modelled and observational SSI data as a solar input. The atmospheric model has been used in many studies to investigate the dynamics and chemical interactions of the Earth's atmosphere (e.g. \citeauthor{BekkiPyle1996}, \citeyear{BekkiPyle1996}, \citeauthor{WarwickBekki2004}, \citeyear{WarwickBekki2004}, \citeauthor{HaighWinning2010}, \citeyear{HaighWinning2010}). It calculates the zonal mean temperature and winds and chemical-constituent concentration in a time-dependent 2D model with full \\radiative-chemical-dynamical coupling. The model considers a spherical Earth with seasons, but without land topography or oceans. The model is resolved into 19 latitudes and 29 pressure levels up to 95 km. In this study the only difference between model runs is the specification of the input SSI. Use of different input SSI affects photochemical reactions involved in ozone production and destruction which account for the largest contribution to heating in the stratosphere, through the absorption of solar flux below 310 nm. The spectral resolution of the input spectra for the atmospheric model decreases non-linearly from less than 1 nm at 116 nm to 5 nm at 300 nm; it remains at 5 nm resolution up to 650 nm, after which it has 10 nm resolution up to 730 nm.

The atmospheric model requires input SSI up to 730 nm. As SORCE/SOLSTICE data are not available above 310 nm, SATIRE-S fluxes are used. This is valid as there is little effect on $\Delta$O$_{3}$ profiles from wavelengths above 310 nm. SORCE/SOLSTICE data show anomalous, large inverse trends in $\Delta$F between 298 and 310 nm, indicated by the dotted lines in Fig.~\ref{fig8}, but tests show that this narrow region also has very little impact on the magnitude and spatial distribution of $\Delta$O$_{3}$; the results remain effectively unchanged if we use SATIRE-S at and above 298 nm, so we use the full wavelength range of SORCE/SOLSTICE in runs using SOLSTICE.

Tests show that the background spectrum (absolute flux) upon which the changes in SSI are superimposed makes very little difference to the  O$_{3}$ results which depend more critically on the spectral shape of the change. This is because of the strong wavelength dependence of ozone production/destruction reactions. We therefore make no attempt to adjust fluxes to a common absolute level and we use the SSI data as published (see also section d of the supplementary material).

Using the same model that we use here, \cite{HaighWinning2010} investigated middle atmosphere ozone changes for SSI changes between 2004 and 2007 for NRLSSI and hybrid SORCE spectra with SIM data for $\lambda >$ 200 nm and an older version of SOLSTICE below. \cite{MerkelHarder2011} presented similar work using the WACCM model, but switched between an older version of SORCE/SOLSTICE and SORCE/\\SIM datasets at 240 nm. The choice of wavelength at which to make the switch is somewhat arbitrary and makes a direct comparison with the runs we do here difficult. Also, the exact choice of dates over which SSI are averaged is different between \cite{HaighWinning2010} and \cite{MerkelHarder2011}. We do not try to reproduce their model setups exactly. Instead, we focus on making comparisons between the models and observations we present here. However, it is worth noting that the spatial distributions of $\Delta$O$_{3}$ when using version 10 of SORCE/SOLSTICE show similar structure and magnitude to the results presented by \cite{HaighWinning2010} and \cite{MerkelHarder2011}. Both studies, when employing SORCE data between 2004 and 2007, found a negative response (i.e. an increasing ozone out-of-phase with solar irradiance) in the mesosphere around 55 km, of 1.2\% and 2\% respectively. They also found a positive response in the middle stratosphere below 40 km, reaching $\sim$2\%. Using NRLSSI in these studies resulted in a response in-phase with SSI changes at all altitudes between 30 and 60 km of up to $\sim$1\%.

\begin{figure*}[t]
 \noindent\includegraphics[width=43pc, angle=0]{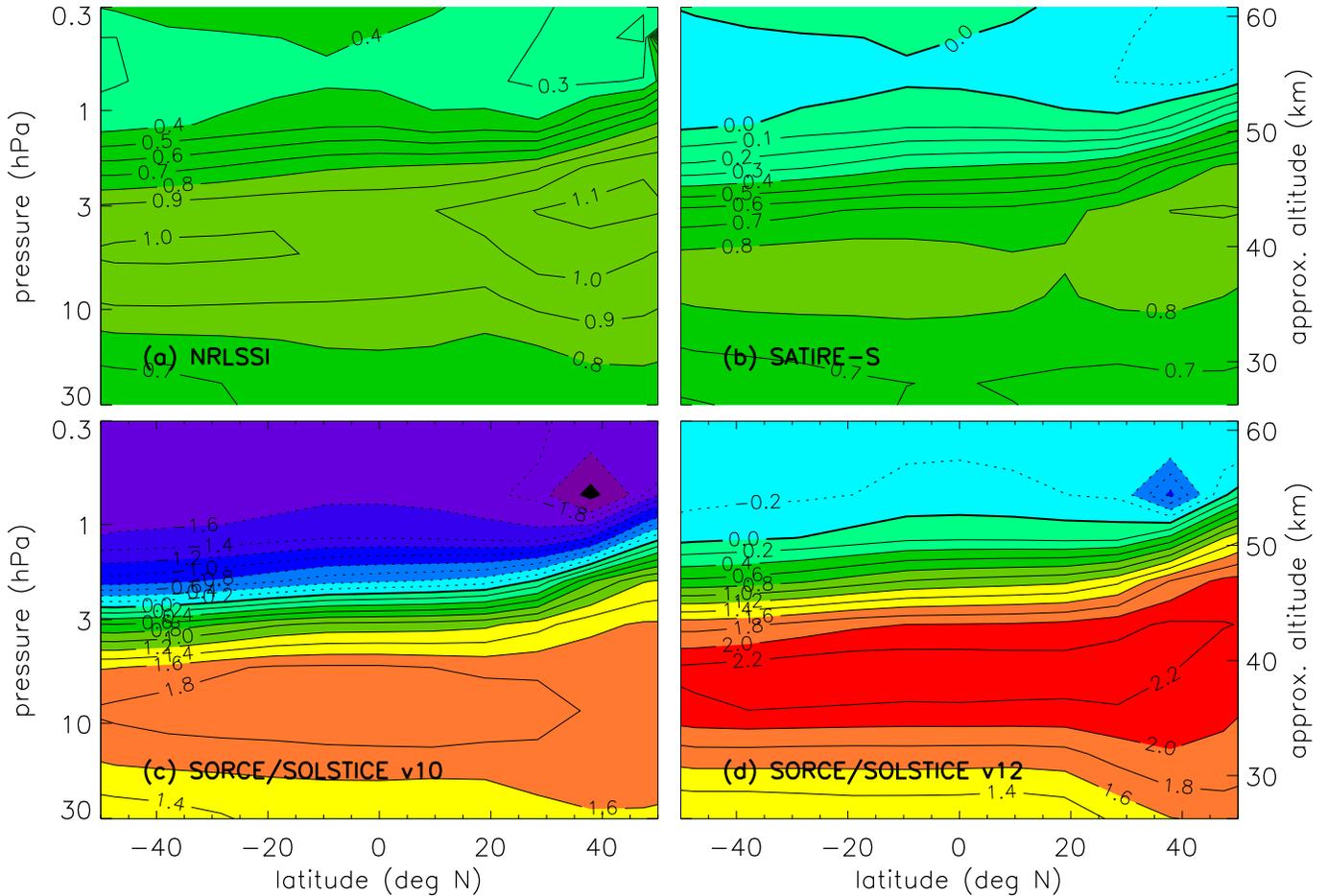}
\caption{Contour plots of the change in stratospheric ozone, $\Delta$O$_{3}$, between 2003 and 2008 using SSI input from (a) the NRLSSI model, (b) the SATIRE-S model, (c) SORCE/SOLSTICE version 10 and (d) version 12. All results are for December 25. Contours are given in 0.1\% levels in (a) and (b) and 0.2\% in (c) and (d) with negative changes, i.e. increases with decreasing solar activity, are shown with dotted contours. Blue and purple reflect negative changes, while green, yellow and red show positive changes.}
\label{fig3}
\end{figure*}

The input UV SSI for the atmospheric model are the 81-day average spectra as presented in Fig.~\ref{fig8}. In Fig.~\ref{fig3} the change in ozone concentration produced by taking the difference between atmospheric model outputs that use SSI from 2003 and 2008 is shown for (a) NRLSSI, (b) SATIRE-S, (c) SORCE/SOLSTICE v10 and (d) SORCE/SOLSTICE v12. The period considered is during a decline in TSI and UV fluxes that is approximately 50\% larger than for the 2004 to 2007 periods used by \cite{HaighWinning2010} and \cite{MerkelHarder2011}. In Fig.~\ref{fig3}a the result for NRLSSI is qualitatively similar in spatial distribution to the NRLSSI result of \cite{HaighWinning2010}, but values are approximately 50\% larger, as expected for the larger UV change. Figure~\ref{fig3}c, using SORCE/SOLSTICE version 10 also displays qualitatively similar $\Delta$O$_{3}$ to those using SORCE presented by \cite{HaighWinning2010} and \cite{MerkelHarder2011}. The general spatial distribution of $\Delta$O$_{3}$ in both versions of SORCE/\\SOLSTICE presented in Figs.~\ref{fig3}c and \ref{fig3}d are similar. However, there are two striking differences: the magnitude of $\Delta$O$_{3}$ in the equatorial lower mesosphere is reduced by a factor of six at around 55 km, from -1.6\% to -0.2\%, for version 10 and 12, respectively; and the zero-line has also shifted up by $\sim$5 km in version 12 with an increased $\Delta$O$_{3}$ maximum in the stratosphere at around 40 km. The change in $\Delta$O$_{3}$ from version 10 to 12 reflects a general decrease of $\Delta$F at wavelengths $>$242 nm and competing increases and decreases in $\Delta$F at wavelengths $<$242 nm.

Despite showing different responses in the atmospheric model for these different datasets, version 10 and 12 of SORCE/SOLSTICE only differ at a small number of wavelengths by more than its stated uncertainty of 0.5\% per annum. So, even though there are these differences, in the $\Delta$Fs of the two SORCE/SOLSTICE versions, they are within the stated uncertainty. Our work shows that knowledge of the detailed spectra to better than this accuracy is required to place any faith in derived atmospheric effects.

The run using SATIRE-S, in Fig.~\ref{fig3}b, also displays a negative $\Delta$O$_{3}$ above $\sim$50 km, though of lower magnitude than for SORCE/SOLSTICE. Relative to NRLSSI, the negative response in the mesosphere results mainly from the larger $\Delta$F in SATIRE-S above 242 nm. The total concentration of O+O$_{3}$ is similar in SATIRE-S and NRLSSI due to similar $\Delta$F for wavelengths $<$242 nm. UV radiation at wavelengths between 242 and 310 nm photodissociates ozone to produce O($^{1}$D) and O$_{2}$. This has two effects: it reduces the O$_{3}$/O ratio and O($^{1}$D) reacts with H$_{2}$O to produce OH which catalytically destroys O$_{3}$. Both processes tend to decrease the O$_{3}$ concentration at these altitudes. Although NRLSSI and SATIRE-S show similar cycle variability below 242 nm, the reduction in ozone concentration at all altitudes and negative response in the mesosphere of SATIRE-S, with respect to NRLSSI, is a result of the larger flux change in SATIRE-S at wavelengths longer than 242 nm. SORCE/SOLSTICE shows very different flux changes compared to SATIRE-S and NRLSSI, but the similar response in the mesosphere relative to SATIRE-S is caused by competing effects on ozone concentration at all wavelengths.

The modelling study by \cite{SwartzStolarski2012} shows similar responses in O$_{3}$ to the NRLSSI and SORCE spectra as \cite{HaighWinning2010} and \cite{MerkelHarder2011}. Analysis of observational ozone data, e.g. SBUV, SAGE and HALOE from \cite{SoukharevHood2006}, Aura/MLS from \\ \cite{HaighWinning2010} and TIMED/SABER from \cite{MerkelHarder2011}, suggests a range of changes in ozone with uncertainties that span the range of differences shown by the model runs here. It is therefore not possible, with datasets and methods currently available, to make unequivocal statements about the `true' ozone response to solar cycle changes. The absolute calibration is less important than the long-term trends in spectral irradiance, the accuracy of which is limited by the stability of the instrument sensitivity. Accurate cycle trends in spectral irradiance are crucial to have confidence in the solar effect on the Earth. At the moment this is lacking.

\section{Discussion and conclusions}
\label{conclusions}


In this paper, we present SATIRE-S spectral solar irradiances covering fully the last three solar cycles spanning 1974 to 2009. SATIRE-S is a semi-empirical model that assumes all irradiance changes are the result of changes in surface magnetic flux \citep{FliggeSolanki2000, KrivovaSolanki2003}. Data are available for all dates over the period with an accompanying error estimate. 

We compare our new SATIRE-S spectral irradiances with the NRLSSI model and SORCE/\\SOLSTICE observations. NRLSSI and SATIRE-S are in good agreement between $\sim$150--242 nm. Between 242 and 400 nm, the solar cycle change in flux in SATIRE-S is generally $\sim$50\% larger than NRLSSI. Both models display significantly lower cycle variability than SORCE/SOLSTICE at almost all wavelengths between 190 and 290 nm. 

Both models provide consistent, long-term reconstructions of solar irradiance with different cycle variability. At the moment there is insufficient reliable observational spectral irradiance data to identify which model is more accurate (on solar cycle time scales). While NRLSSI uses solar indices, SATIRE-S provides a direct translation from observed solar images to irradiances. The results presented in this paper show that SATIRE-S reproduces the PMOD TSI composite values better than NRLSSI (though the uncertainties of SATIRE-S and PMOD both encompass NRLSSI). Above 250~nm, the SATIRE-S spectral response is closer to that shown by UARS/SUSIM than NRLSSI (see inset of Fig~\ref{fig7}). We suggest that the SATIRE-S spectral irradiances should be considered in future climate studies, either on its own or in addition to NRLSSI.

We present an example of the physical implications that these different SSI datasets have within the stratosphere. We calculate the ozone concentration resulting from the use of the three datasets within a 2D atmospheric model. We find that: (a) the magnitude of mesospheric $\Delta$O$_{3}$ response when using SORCE/SOLSTICE versions 10 and 12 is very different; (b) SATIRE-S mimics the small negative mesospheric solar cycle change that SORCE/SOLSTICE version 12 produces; and (c) NRLSSI displays positive changes at all heights.

Interestingly, a recent study by \cite{WangLi2013} suggests that modelled solar cycle changes in OH concentration in the stratosphere agree better with observations when using SORCE than NRLSSI data as a model input. We note that \cite{WangLi2013} used hybrid SORCE datasets with SIM data either above 240 or above 210 nm and SOLSTICE below these wavelengths; this is different to the spectra we use here or that used by \cite{HaighWinning2010} and \cite{MerkelHarder2011}. However, the different ozone changes in the mesosphere resulting from SSI inputs from two different models and two versions of the same observational dataset in this study, highlights the need for a better understanding of solar cycle changes at wavelengths important for both O$_{2}$, O$_{3}$ and other atmospheric photochemistry such as OH as in \cite{WangLi2013}. If the solar effect on ozone is to be isolated correctly, it is imperative that greater certainty is established in SSI variability.

This raises an important point that should be considered when making comparisons of observed with modelled ozone changes which use different SSI datasets. Data are usually updated to make an improvement on previous releases. Unfortunately, the fast pace at which data revisions are made makes comparisons with previous publications difficult. In the case presented here, the change in $\Delta$O$_{3}$ between version 10 and 12 of SORCE/SOLSTICE leads to a result in the mesosphere that is so different that it is not yet possible to make robust conclusions about SSI or ozone based on the negative ozone mesospheric response. Many previous investigations have used different SORCE/SIM spectral data, and hybrids with SORCE/SOLSTICE, to investigate the atmospheric response. This makes direct comparisons with the results published in the literature difficult. We have focused the investigation to highlight what is different about SATIRE-S and to only consider the effect of one SORCE dataset in each run. A major result is that significant uncertainties remain in how SSI affects the atmosphere on time scales longer than the solar rotation.

The SATIRE-S SSI data are available for download at http://www.mps.mpg.de/projects/sun-climate/data.html. \\The SSI reconstruction ends in October 2009. Work is currently underway to extend the reconstruction using full-disk images from the Solar Dynamics Observatory Helioseismic and Magnetic Imager \citep{SchouScherrer2012} and will be made available for download in the future.

\begin{acknowledgment} 
We thank Marty Snow, Linton Floyd, Tom Woods, Peter Pilewskie and Judith Lean for helpful discussions. We acknowledge the use of the Ottawa/Penticton 2800 MHz Solar Radio Flux and version d41\_62\_1003 of the PMOD dataset from PMOD/WRC, Davos, Switzerland and the unpublished data from the VIRGO Experiment on SoHO, a project of international cooperation between ESA and NASA. This work has been partly supported by the NERC SolCli consortium grant, the STFC grant ST/I001972/1, the WCU grant number R31-10016 funded by the Korean Ministry of Education, Science and Technology and FP7 SOLID.
\end{acknowledgment}

\ifthenelse{\boolean{dc}}
{}
{\clearpage}
\bibliographystyle{ametsoc}
\bibliography{BallKrivova_JAS_arxiv_Aug2014}



\ifthenelse{\boolean{dc}}
{}
{\clearpage}
\pagebreak
-
\pagebreak
\section*{\begin{center}Supplementary Materials\end{center}}

\setcounter{figure}{0}
\makeatletter 
\renewcommand{\thefigure}{S\@arabic\c@figure}


\subsection{Cycle variability}
\label{supplyman}

\begin{figure*}[t]
 \noindent\includegraphics[width=31pc, bb=18 43 563 797, angle=90]{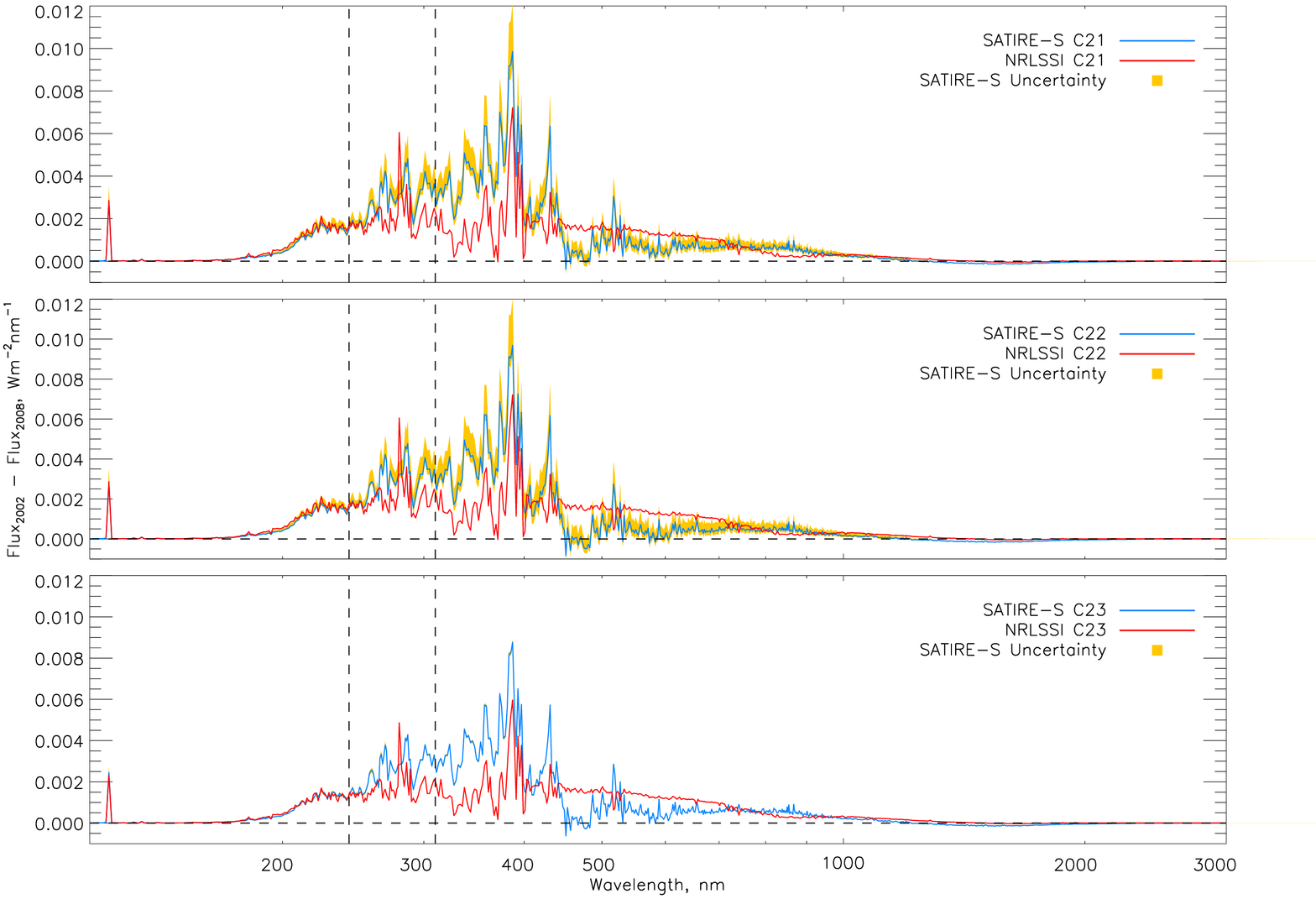}
\caption{As for the lower panel of Fig.~1 of the main paper, but with cycle amplitude uncertainty included as yellow shading for SATIRE-S. This is given for cycles 21 (top), 22 (middle) and 23 (bottom). Cycle maximum and minimum dates are taken from the monthly sunspot numbers. Cycle 23 in this plot is for the 81-day average fluxes between March 2000 and December 2008, whereas Fig.~1 is between February 2002 and December 2008, which had a higher mean irradiance than around March 2000.}
\label{figs1}
\end{figure*}

The uncertainty of SATIRE-S is time-dependent and increases going backwards in time from 2008. This is due to the way the reconstructions from the different instruments are combined (see \citeauthor{BallUnruh2012}, \citeyear{BallUnruh2012}, and section~2a of the main paper). We show the estimated cycle uncertainty, as a function of wavelength, for three solar cycles in Fig.~\ref{figs1}. The bottom panel of Fig.~\ref{figs1} is similar to the bottom panel of Fig.~1 in the main paper, but is now between March 2000--December 2008 and has the 1$\sigma$ uncertainty estimate included as yellow shading. The uncertainty is very small in cycle 23 and is difficult to see. However, cycles 21 (top; December 1979--September 1986) and 22 (middle; July 1989--May 1996) show much larger uncertainty due to steps (ii) and (iii) of the construction method explained in section~2a. We note that even with the larger uncertainty in cycles 21 and 22, it does not generally encompass the cycle changes of NRLSSI for which there is no uncertainty estimate.

\subsection{Gap filling}
\label{suppgap}

\begin{figure*}[t]
\noindent\includegraphics[width=43pc, bb=0 4 569 215]{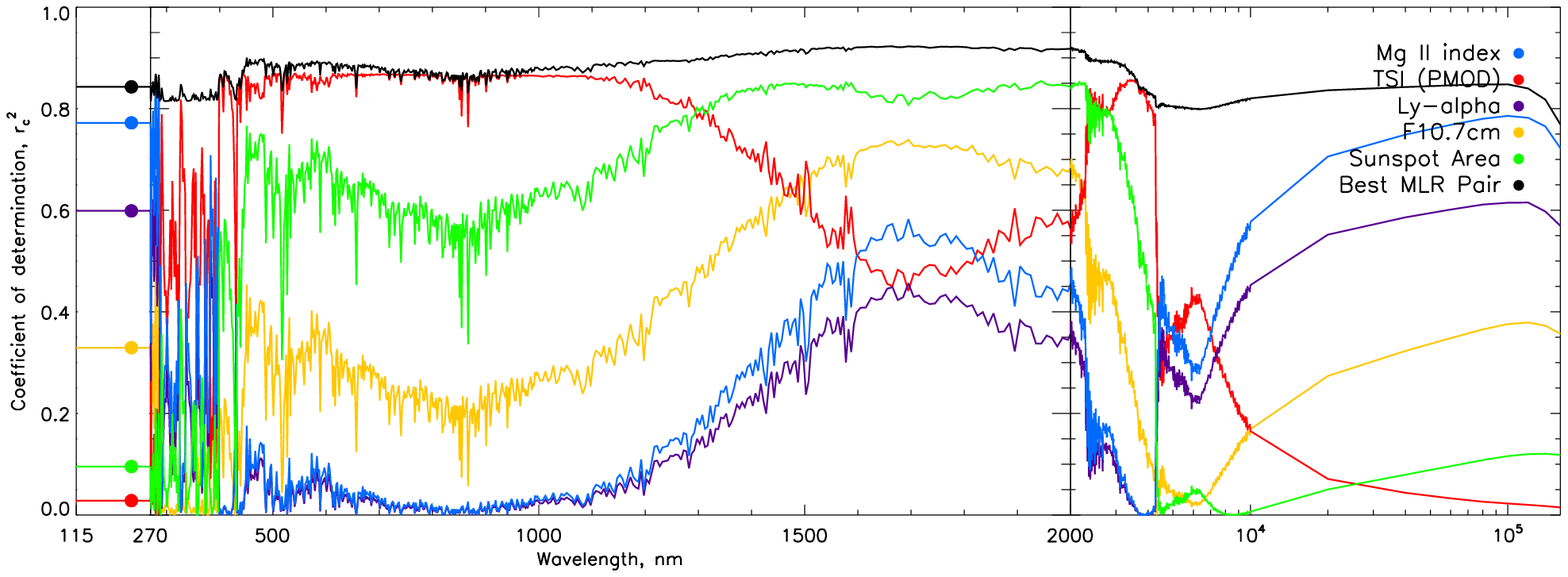}
\caption{The coefficient of determination, r$^{2}_{c}$, between the detrended, rotational variability of each wavelength in SATIRE-S and five solar indices: the Mg II index (blue), the PMOD TSI composite (red), the Lyman-$\alpha$ composite (purple), the F10.7 cm radio flux (yellow) and the sunspot area (green). Also shown, in black, is the highest value of r$^{2}_{c}$ from the multi-linear regression of all pairs of the indices.}
\label{figs8}
\end{figure*}

We use five solar indices to fill gaps in the SATIRE-S time series. Here we detail the behaviour of these indices and show examples of the gap-filled time series in three spectral bands.

Each index is indicative of the behaviour of some feature (or combination of features) or source of radiative flux in the solar atmosphere, which in turn should correlate with activity and therefore variability in solar spectral irradiance. The variability of each index depends on the time scale being considered. We only use the detrended indices to avoid biasing the long-term SATIRE-S SSI behaviour to any index.

To account for the rotational variability and fill gaps in the detrended SSI time series, we compare each detrended index with each detrended SATIRE-S SSI wavelength. We do this for all wavelengths above 270 nm and for the integrated region between 220--240 nm. Below 270 nm, all time series are determined from the integrated 220--240 nm region so only this one time series needs to be filled to account for all wavelengths in this wavelength region (see section~2b of main paper). In Fig.~\ref{figs8}, the coefficient of determination, r$^{2}_{c}$ (the square of the correlation coefficient, r$_{c}$), shows how well each detrended index can account for the rotational variability at each wavelength. Colours are given in the legend. Also shown is the highest value r$^{2}_{c}$ from the multi-linear regressions of all pairings of indices (black). The scale on the x-axis is linear between 115 and 2000 nm and logarithmic above. The constant value of r$^{2}_{c}$ below 270 nm reflects the use of the 220--240 nm region, where the dots are located, to determine the wavelengths in this region.

Considering just the individual indices, we briefly describe the source of their variations (except for TSI), and why they agree with the SSI variability in various wavelength intervals. Below 300 nm, the Mg II index agrees best with SATIRE-S SSI wavelengths, accounting for around 70--80\% of the rotational variability. The Mg II index is a measure of the ratio of the emission reversal in the line core of the Mg II line (at $\sim$280 nm) and the Mg II wings. The line core is formed higher up in the solar atmosphere than the wings and is associated with regions of strong magnetic flux, as found in faculae and plages. Faculae dominate spectral variability below 300 nm, so the agreement of the Mg II index with these wavelengths is to be expected.

\begin{figure*}[t]
\noindent\includegraphics[width=55pc, bb=10 325 707 750]{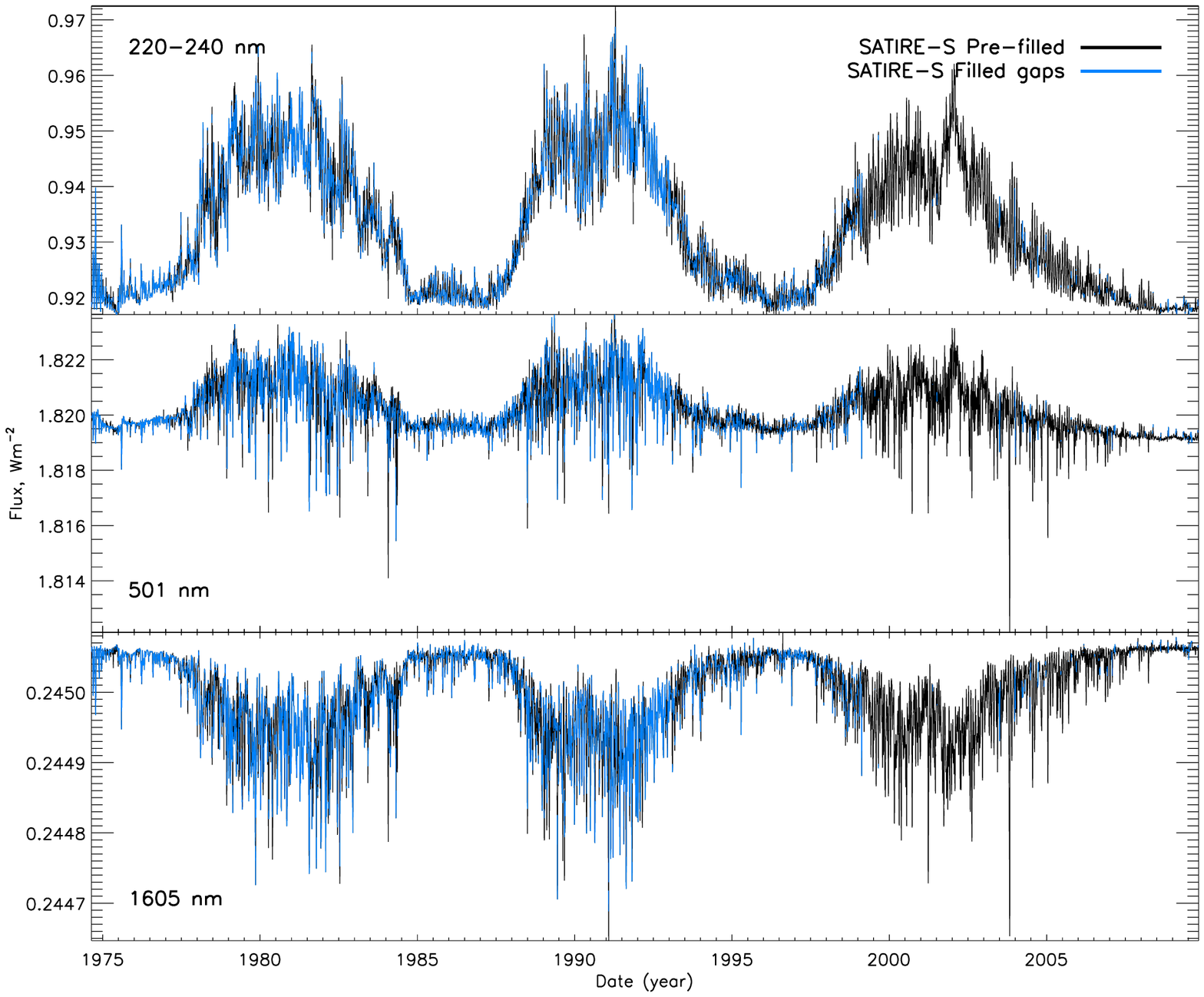}
\caption{Time series for three wavelength regions from SATIRE-S showing both the reconstruction direct from the model (black) and the gaps in the data that have been filled with proxy data (blue). The top plot shows the integrated 220--240 nm flux, the middle plot is for the 2 nm band at 501 nm and the bottom plot is the 10 nm band at 1605}
\label{figs3}
\end{figure*}

Between 300 and 400 nm, there is a transition between the dominance of faculae and of sunspots on spectral variability. This is reflected in the behaviour of the indices over this spectral interval where correlation coefficients are highly variable and quite poor at some wavelengths. The lower rotational agreement is not so much of a concern for cycle-length changes in these wavelengths, which is more important in the analysis of the long-term impact of SSI on, e.g., stratospheric chemistry. By combining two proxies, which represent sunspot and facular variations, high agreement ($>$80\%) with SATIRE-S wavelengths in this region can be maintained, as shown with the black line in Fig.~\ref{figs8}.

The regions between 400 and 1300 nm and 2300 and 4000 nm are best represented by TSI, which can represent more than 80\% of the variability at almost all of these wavelengths. TSI is replaced by the sunspot area (SSA) and F10.7 cm radio flux between 1400 and 2200 nm as being the best representative single index of rotational variability; this region contains the opacity minimum and represents the deepest directly observable layers of the photosphere where sunspots and pores dominate and where faculae are expected to appear dark. Where the transition to dark faculae occurs is dependent on the facular model atmosphere \citep{FontenlaAvrett1993, UnruhSolanki1999, UnruhKrivova2008}.  As the TSI includes facular brightening, its correlation with the near-IR SSI decreases. We thus find that the SSI is best represented by the SSA and F10.7 cm radio flux. The F10.7 cm radio flux is formed from plasma trapped in coronal loops anchored to sunspots, so, although it is not formed in the photosphere, the F10.7 cm radio flux is highly correlated with the SSA on rotational time scales and hence the modelled irradiance at these wavelengths. Between 4000 and 10 000 nm none of the indices agree well with rotational variability in SATIRE-S SSI. As in the 300-400 nm region, there is a transition between a stronger influence from pores and sunspots below 4500 nm and faculae above; the contrast of faculae increases monotonically from here while penumbral and umbral contrasts remain relatively constant. Once again, by combining two indices, good agreement with the SATIRE-S rotational variability is found.

In Fig.~\ref{figs3}, we give three examples of the completed gap-filling process. The integrated region of 220--240 nm and the spectral regions centred at 501 (at 2 nm resolution) and 1605 nm (10 nm) are plotted with the original SATIRE-S time series (black) and the additional daily gap-filled data (blue). The time series is consistent on long- and short-time scales.

\subsection{Uncertainties in model spectrum}
\label{suppmodel}

\begin{figure*}[t]
\noindent\includegraphics[width=30pc, bb=7 5 567 815, angle=90]{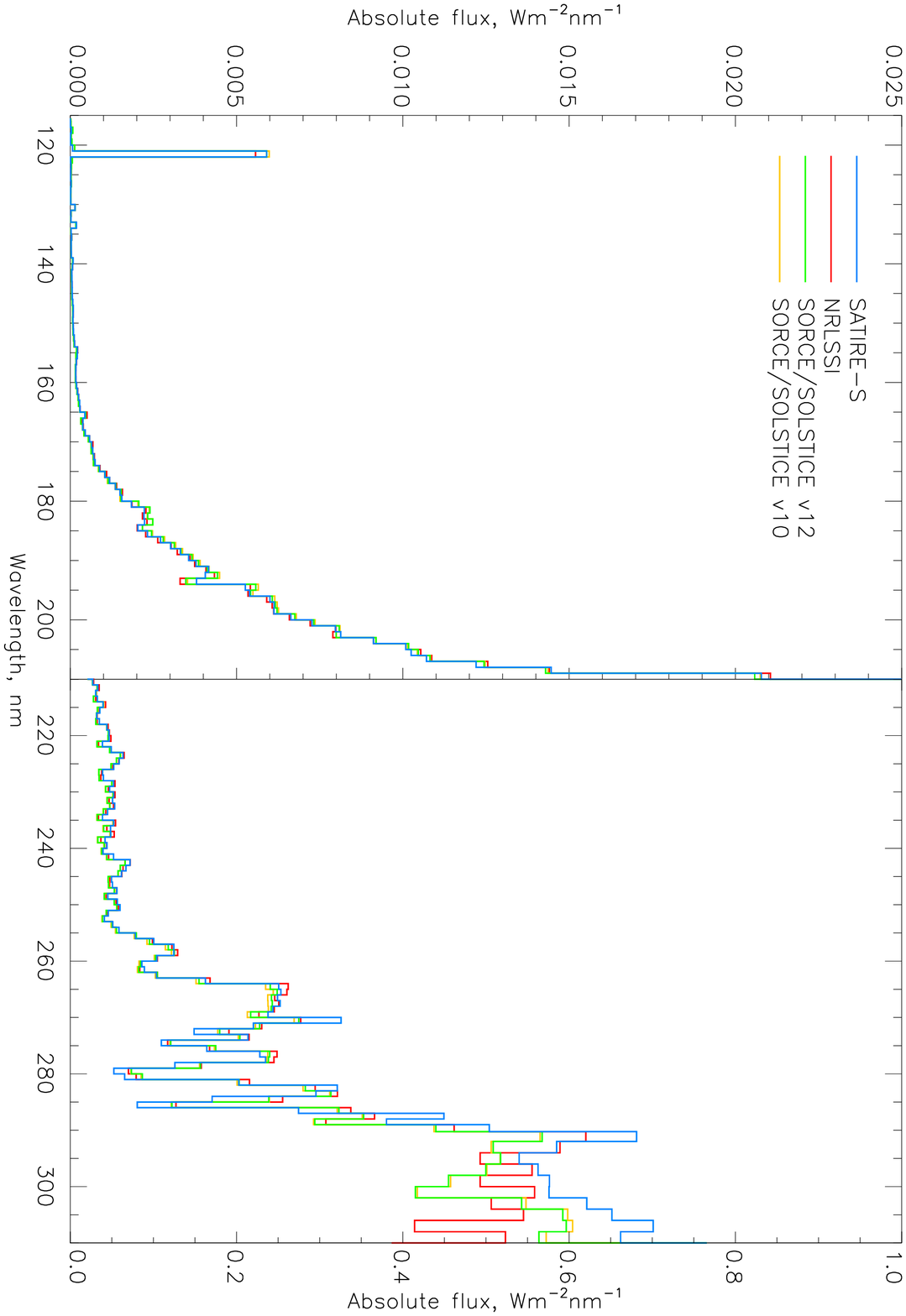}
\caption{The absolute flux for the 81-day period centred on the solar minimum of December 2008 for SATIRE-S (blue), NRLSSI (red), SORCE/SOLSTICE v12 (yellow) and SORCE/SOLSTICE v10 (green).}
\label{figs2}
\end{figure*}

The uncertainties we consider in the spectral reconstruction do not include an uncertainty in the model atmospheres themselves, which are estimates of the structure of the solar atmosphere and are constrained using spatially resolved observations. Some of the differences between NRLSSI and SATIRE-S at wavelengths above 400 nm are due to the use of model atmospheres with slightly different atmospheric profiles. For example, the extent of the near-IR variability, and even its sign (correlated or anti-correlated with TSI), depends very sensitively on the adopted facular model. We note that the intensities are derived from one-dimensional plane parallel atmospheres and thus do not take the geometry of magnetic features into account.

Another consideration is the assumption of local thermodynamic equilibrium (LTE) conditions in the solar atmosphere when employing the ATLAS9 code to calculate model intensities. This leads to errors in the modelled irradiance variability in some wavelength regions, mainly in the UV. In the data release, we flag the wavelength regions where the uncertainty due to the use of LTE is large, such as, e.g., Mg I line at 285~nm. In Fig.~8a of \citep{UnruhKrivova2008}, the variability of SATIRE-S and SORCE/SIM wavelengths between 220 and 290 nm is shown. Above 270 nm, only the Mg I line in SATIRE-S has much larger rotational variability than SORCE/SIM (more than four times); other lines showing much larger rotational variability in the UV below 270 nm are corrected using the method described in \citep{KrivovaSolanki2006} and section~2a of this paper. For wavelengths longer than 285 nm, we flag those that show rotational variability exceeding SORCE/SIM or the instrument noise by 25\%. This includes the Ca II H and K lines and the wavelength region around 385 nm (see Fig.~8b of \citeauthor{UnruhKrivova2008}, \citeyear{UnruhKrivova2008}).

\begin{figure*}[t]
\noindent\includegraphics[width=52pc, bb=10 325 707 750]{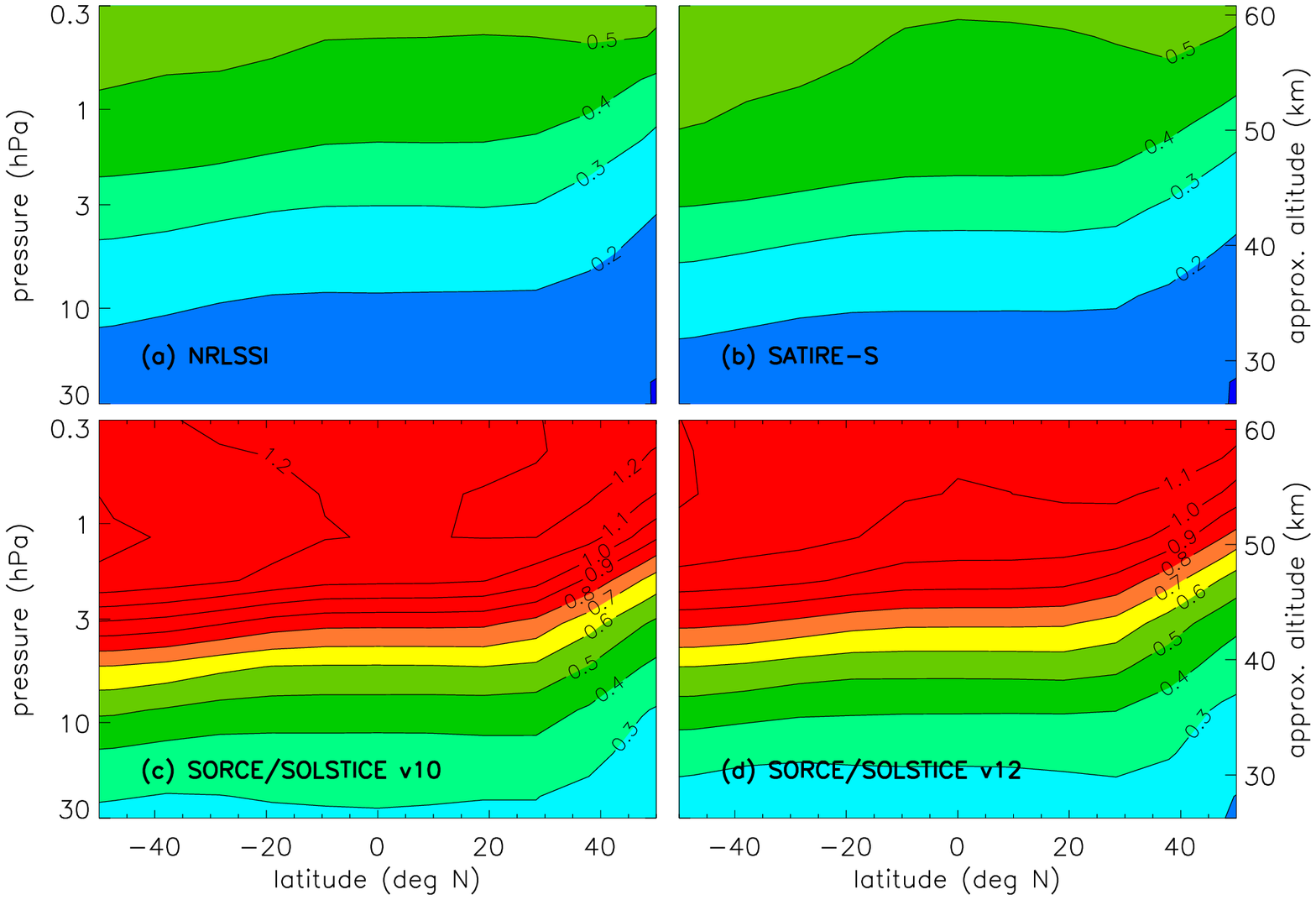}
\caption{The same as for Fig.~\ref{fig3} of the main paper, but showing the change in stratospheric temperature. Contours are given in 0.1 K levels.}
\label{figs5}
\end{figure*}

\begin{figure*}
\noindent\includegraphics[width=43pc, bb=9 15 593 587]{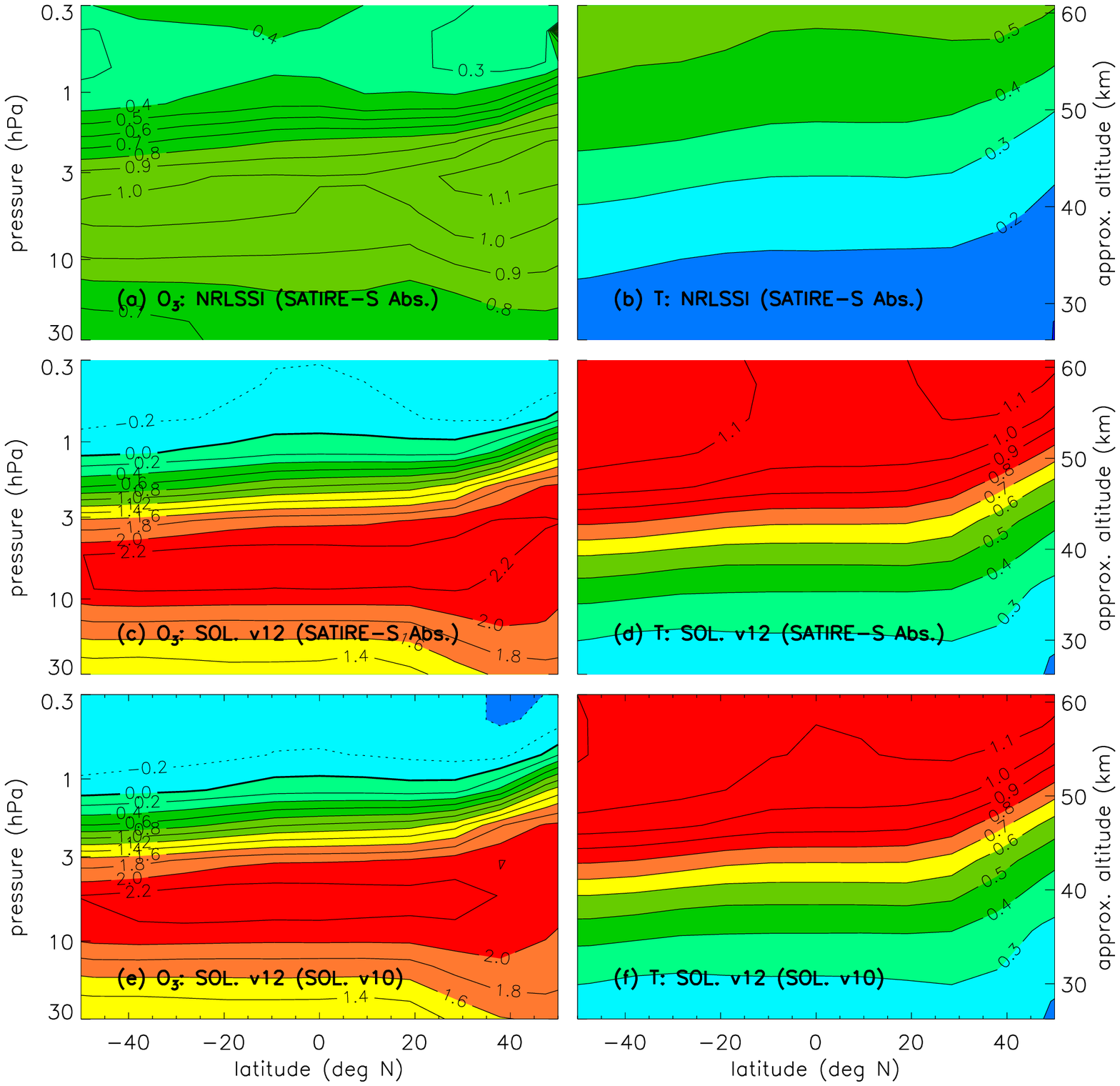}
\caption{As for Fig.~\ref{fig3} of the main paper and Fig.~\ref{figs5}, contour plots of the change in stratospheric ozone (left panels) and the corresponding temperature change (right panels). The top row shows ozone and temperature changes when employing NRLSSI in the HP model with the absolute flux level of SATIRE-S. The middle row is using SORCE/SOLSTICE v12 change in flux with the absolute flux level of SATIRE-S. The bottom row is using SORCE/SOLSTICE v12 change in flux with the absolute flux of SORCE/SOLSTICE v10.}
\label{figs6}
\end{figure*}

\subsection{Effect of absolute flux on ozone and temperature}
\label{supptemp}

We show here that the choice of absolute flux has very little impact on both the cycle change in ozone and temperature in the stratosphere and lower mesosphere. We do this to support the claim made in section~4 of the main paper that it is the relative change between cycle maximum and minimum that is more important when considering cycle changes in ozone. Figure~\ref{figs2} gives the mean absolute fluxes of SATIRE-S (blue), NRLSSI (red) and SORCE/SOLSTICE versions 12 (green) and 10 (yellow) for the three month period centred on December 2008. The left y-axis scale corresponds to wavelengths below 210 nm and the right scale is for wavelengths longer than this. We see that all four datasets show a difference in absolute flux, but the largest differences are above 290 nm, where relative cycle variability is lowest in the models. SATIRE-S has the largest absolute flux at these wavelengths, while SORCE/SOLSTICE (at 290--300 nm) and NRLSSI (at 300-310 nm) show the lowest.

Figure~\ref{figs5} gives the change in temperature, in degrees Kelvin, between 2003 and 2008, corresponding to the same runs that produced Fig.~\ref{fig3} in the main paper. All plots in Fig.~\ref{figs5} show an increasing change in temperature with higher altitude, though NRLSSI and SATIRE-S both show about half the change in temperature compared to SORCE/\\SOLSTICE. SORCE/SOLSTICE version 10 shows a slightly higher change in temperature than version 12 at all altitudes.

In Fig.~\ref{figs6}, we show that the choice of absolute flux has little impact on the change in temperature and ozone concentration. We demonstrate this for three cases: (i) by putting the $\Delta$F between 2003 and 2008 from NRLSSI on the absolute flux of SATIRE-S with the resulting ozone and temperature response given in Fig.~\ref{figs6}a and b, respectively; (ii) by using the $\Delta$F from SORCE/SOLSTICE v12 with the absolute flux of SATIRE-S in Fig.~\ref{figs6}c and d; and (iii) the same as (ii), but with the absolute flux of SORCE/SOLSTICE version 10 in Fig.~\ref{figs6}e and f. Comparing Fig.~\ref{fig3}a of the main paper with Fig.~\ref{figs6}a (for ozone concentration) and Fig.~\ref{figs5}a with Fig.~\ref{figs6}b (for temperature), it can be seen that the change resulting from using either the absolute flux of NRLSSI or SATIRE-S does make a small change to the contours, but these changes are much less than the difference between contour levels. Minor differences are more easily detectable for the examples using SORCE/SOLSTICE v12 changes with the absolute fluxes of SORCE/SOLSTICE v12 (Figs.~\ref{fig3}d of the main paper and ~\ref{figs5}d), SATIRE-S (Figs.~\ref{figs6}c and d) and SORCE/SOLSTICE v10 (Figs.~\ref{figs6}e and f). However, once again, these differences are very small and only stand out at the winter pole at 0.3 hPa. It is, therefore, the case that the relative variations in the solar spectral irradiance have a much more significant effect on changes in ozone concentration and temperature in the stratosphere than the choice of absolute flux, at least for the range of absolute fluxes from datasets considered in this study.

\end{document}